# Optimal Scheduling for Maximizing Information Freshness & System Performance in Industrial Cyber-Physical Systems

Devarpita Sinha, *Graduate Student Member, IEEE* and Rajarshi Roy, *Senior Member, IEEE*

*Abstract--"Age of Information"* is a newly introduced metric, getting vivid attention for measuring the freshness of information in real-time networks. This parameter has evolved to guarantee the reception of timely information from the latest status update, received by a user from any real-time application. In this paper, we study a centralized, closed-loop, networked controlled industrial wireless sensor-actuator network for cyber-physical production systems. Here, we jointly address the problem of transmission scheduling of sensor updates and the restoration of an information flow-line after any real-time update having hard-deadline drops from it, resulting a break in the loop. Unlike existing real-time scheduling policies that only ensure timely updates, this work aims to accomplish both the time-sensitivity and data freshness in new and regenerative real-time updates in terms of the 'age' of information. Here, the coexistence of both cyber and physical units and their individual requirements for providing the quality of service to the system, as a whole, seems to be one of the major challenges to handle. In this work, minimization of staleness of the time-critical updates to extract maximum utilization out of its information content and its effects on other network performances are thoroughly investigated. A greedy scheduling policy called *"Deadline-aware highest latency first"* has been used to solve this problem; its performance optimality is proved analytically. Finally, our claim is validated by comparing the results obtained by our algorithm with those of other popular scheduling policies through extensive simulations.

## I. INTRODUCTION

**I**N In the era of machine to machine (M2M) communication, Internet of Things (IoT), and Cyber-Physical systems (CPSs) [1] numerous real-time systems need to communicate and coordinate with each other by exchanging a large amount of time-varying information in a secure and timely manner (Ultra-Reliable Low Latency Communication-URLLC) through next-generation (5G & beyond) wireless communication networks.

Age of Information (AoI) [2] has emerged as one of the main focus of interest for the time-critical status updates in real-time applications such as sensor updates or control commands for actuation tasks in CPSs or IoT for remote health monitoring, smart home, factory or smart transportation system, etc. Here time-stamped status updates are generated and transmitted as packets containing updated information about one or more variables of interest of the source [3]. If $u(t)$ be the random generation instant of the last packet received by the destination, the random process that defines the 'age' at any instant $t$ is $\Delta(t) \triangleq t - u(t)$ [2-3]. Age minimization for ensuring high information freshness can be achieved by delivering packets at regular intervals with a delay as minimum as possible [4].

### A. Background and Motivation

Our work focuses on maintaining information freshness and timeliness during the information exchange within the industrial cyber-physical systems (ICPSs). This work is motivated by the unavoidable requirement of monitoring individual plants in order to maintain them at their stable operating point in an industrial factory automation system. Here, for monitoring purpose, the updated status about the state evolution of the underlying physical processes are sensed by the sensors as an event-triggered or time-triggered manner and they are transmitted to the controller through a wireless mesh network. Upon receipt of those sample packets within their deadlines, controllers are responsible for the decision making and sending those appropriate control instructions to the corresponding actuators. However, the data processing center within the controller either performs a data fusion using consecutive samples of the same monitoring point or it may use an individual sample for decision making. Actuators, then, control the underlying physical processes according to the instructions received from the controller based on the current state of the system. In this way, a closed-loop wireless sensor-actuator network (WSAN), as shown in Fig 1, is formed by enabling the underlying system to sense, interact, and duly change their physical behaviors in real-time by integrating the cyber and physical worlds [5]. Some of the issues involved in this closed loop control process are as follows:

a.  *What is the best possible mode of sampling? or, simply in other words: (i) How to sample? [6-8] and (ii) When to sample? [2, 9-10]*
b.  *What parameters need to be sampled?*
c.  *How sampling through queues affect the situation? [11]*
d.  *How many updates should be used for decision making? [12-13]*
e.  *How the decision or estimation is computed? [14-15]*
f.  *What are supposed to be the cost functions for handling a corresponding optimization problem? [16]*
g.  *What actions should be taken if the update or control packet is not received due to packet loss or error in channel? [17-18]*
h.  *How to handle limited network resources? [19]*
i.  *How to take preventive measures against security threats? [20], etc.*

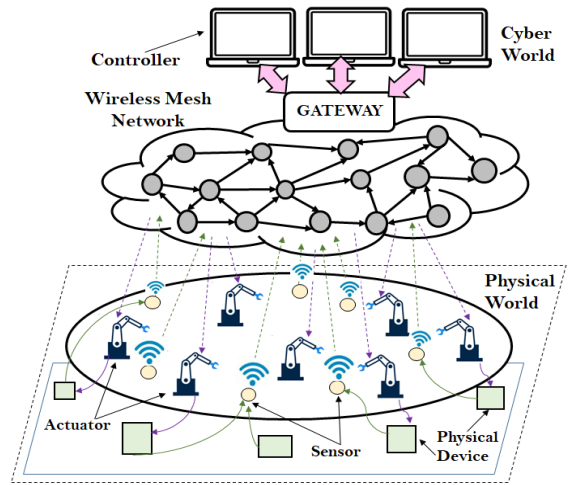

Fig. 1. Closed loop WSAN for in ICPS.

A scheduling [21] among the updates from different flow-lines is required if the uplink bandwidth is a curse. While doing the same, as we are concerned about the real-time mitigation of the problems, the plant is facing the need to schedule those updates before they violate their corresponding deadline. In this paper, AoI is considered as the determinant of scheduling the unattended active sensor updates for processing with an aim to attain maximum average utility value out of those information. AoI is termed as the amount of time, passed after the generation of the sensor sample that has most recently processed by the controller. Here 'age' of a sensor update gets primary importance as it can capture the 'age' of the status of the phenomenon under consideration that plays a vital role, indeed, in maintaining the stability of the system.

Minimizing AoI can be viewed equivalently as an attempt to minimize the staleness or to maximize the value of information. Therefore, if a new update from the same flow appears before the processing of previous update, then the new one might be allowed to replace the previous one and the AoI increment process continues. The content of the new update carries the latest status information of the plant which enables the processor to take a more accurate decision. However, not just this value rather the age of the phenomenon under consideration along with this value may be required for appropriate decision making at the data processing center. If the sensor samples fail to be processed within their respective deadlines, they are dropped from the system and the corresponding flow-line (sensor-actuator loop) breaks down. However, failing to exchange real-time information almost instantly and reliably may cause production loss, plant shutdown and/or fatal accidents in smart factories [22].

Bad channel conditions, communication delays, and congestions due to limited availability of resources make the situation very challenging to meet the maximum freshness of information under stringent deadline constraint. Therefore, advanced research on deadline-aware real-time scheduling for optimizing information freshness and other network performances is on high priority. WirelessHART is one of the most popular standards for industrial wireless sensor-actuator networks (IWSAN) on top of IEEE 802.15.4 [23]. This standard adopts the time division multiple access (TDMA) technique in its MAC layer to avoid possible collisions between simultaneous transmissions through a common wireless channel [24].

In this paper, we provide a deep insight into finding an offline, work conserving, energy-efficient, dynamic priority-based deadline-aware sensor scheduling algorithm that will help in maximizing the freshness of information in addition to the minimization of the number of packet drops. However, if unfortunately, packet drops under any circumstances, our paper tries to find a solution, too, to re-establish the broken flow-line.

*B. Related Literature*

At present, different types of research are going on in this direction. Works done in references [25-27] simply analyze AoI for packets with different deadlines but do not find any scheduling scheme to minimize it. Paper [25] provides a simulative study about the impact of buffer sizes, packet deadlines, and packet replacement in calculating the average age of information for different queuing systems. Whereas, paper [26] considers the effect of random and exponential packet deadlines as a controlling factor on the average age of information for M/M/1/1 and M/M/1/2 queuing systems and indicates the importance of limiting the maximum deadline. Moreover, it concludes that using a deadline can provide a much better result than both the M/M/1/1 and M/M/1/2 without a deadline. Paper [27] studies AoI for infinite capacity M/G/1+G queues with packet deadline. Another set of researches [28-31] investigates scheduling algorithms to maximize the information freshness in the real-time wireless networks. Reference [28] investigates optimal scheduling strategy for an NP-Hard problem to minimize the overall information age for multi-source, common channel wireless network with an objective to deliver the information as timely as possible. Age optimal scheduling policy preemptive Last Generated First Served (LGFS) is studied in [29] for optimizing the age-of-information without the loss of packets in both infinite and finite buffer queues. Scheduling problems for the average and peak AoI optimality in wireless networks under general interference constraints and time varying channel are considered in [30-31]. Scheduling policies for maintaining information freshness in CPS are presented in [32-34] but they do not consider any deadline for those packets. Paper [32] tries to maintain the freshness and synchrony of information from multiple sensors to improve the effectiveness of the control decisions in a distributed cyber-physical system. Age-based scheduling strategy for the joint optimality of data sampling and link scheduling to gain better information freshness in multi-hop, multi-flow CPS is proposed in [33]. Our previous work in [34] proposes two greedy scheduling policies: one is highest age first (HAF) for expected age minimization and another one is highest latency first (HLF) for minimizing the expected latency involved in the age of sensor samples in an ICPS. This paper can eventually proves that minimum age does not always guarantee the maximum freshness of information and satisfactory system performance at the same time in CPSs. Some of the references [35-36] provide deep insight on packet inter-delivery time optimization in real-time sensor networks or CPSs based on risk factors, and the last but not the least [37] deals with the deadline-aware packet scheduling for age minimization for real-time traffic.

*C. Contribution*

To the best of our knowledge, our previous work [38] is the first one that tries to collectively implement a deadline-aware packet scheduling policy to maximize the information freshness for the real-time control of WSAN in the ICPSs. Here, the cost function (utility of information) is designed in such a way that it captures the effect of both the latency or waiting time (time passed after entering the system) and laxity (remaining time before deadline) of the packet. A scheduling policy deadline-aware highest latency first (HLF-D) is proposed in the paper [38]. This optimizes the above mentioned goal of providing deadline-constrained service to the status updates in such a way that the overall quality of service (QoS) in the system performance (combining both in cyber and physical domain) is preserved.

But the limitation of the paper lies in its consideration that if a packet, unfortunately, misses its deadline, the flow-line is going to stop service forever and the corresponding actuator gets a signal to take some preset default control action (or shut down, eventually). It cannot accommodate any possibility of future restoration of the broken flow-lines. However, any interruption in flow-line may cause significant production loss or provide extra load on remaining flow-lines in the network.

However, in this paper, we try to find solutions for bringing the broken flow-lines back to action. Here, along with the objective of finding a deadline-aware scheduling policy in order to maximize the information freshness in addition to minimum packet loss, we also handle the case of restoration or replacement of the broken flow-lines, after some maintenance work, from which packets have been dropped some time before. The main contributions of this paper are listed below:

- *Tries to maximize the freshness of information while maintaining the number of packet drops as minimum as possible. For doing this, a new quality metric "Utility of information" is introduced.*

- *Incorporated the regenerations of broken flow-lines (flow-lines from which packets have been dropped) once their repairing is done. Packets from these sensors join the scheduling process again and compete with other sensor samples for getting service.*

- *While formulating the utility value of a sensor sample, apart from its freshness and laxity, we also consider the number of times its corresponding flowline already attempted regenerations.*

- *Introduced a penalty function affecting the utility of the tardy packets in the system.*

- *Applied Deadline-aware Highest Latency First (HLF-D) algorithm, proposed in our previous work [38], for this system model, too, and mathematically proved that this algorithm gives the global optimal solution.*

- *Compared the performance of HLF-D algorithm with other popular algorithms in providing the quality of services (QoS) in terms of utility, age, latency and RMS jitter for our system model and validate our claim by simulation results too.*

These make our work distinct from [38] and other existing literature. This kind of problem is relevant for applications like cyber physical production system (CPPS) [39] for smart manufacturing under Industry 4.0.

### D. Organization

The rest of the paper is organized as follows. Sec II describes the system model and sec III derives important mathematical formulations related to it. In sec IV, a greedy sensor scheduling algorithm HLF-D is discussed and its optimality is proved analytically in terms of the utility of information content present in the fresh and regenerated sensor samples. Simulation results and discussions are given in sec V followed by the conclusion of this paper in sec VI.

### E. Notation

In this paper, before the term '*attempt number*' is introduced, age, deadline, and reset time are represented as simple $h_{t,i}(k_i)$, $D_i(k_i)$ and $RS_i(k_i)$, respectively. After the introduction of attempt number $\xi_i$ of $i^{th}$ flow-line, any attribute of the sensor sample, represented with the notation $[.]_{t,i}(k_i)$, indicates that it is the total value of that parameter incurred up to the time slot $t$ after the first arrival of $k_i^{th}$ sample from sensor $i$. On the other hand, any parameter with the notation $[.]_{t,i}^{\xi_i}(k_i)$ represents the value of that parameter for $k_i^{th}$ sample from sensor $i$ at any slot $t$ during its $\xi_i^{th}$ attempt to transmit. Any notation in the form $[.]_i(k_i)$ presents a parameter of the sample $k_i$ from $i^{th}$ flow-line which is independent of time as well as attempts. Whereas, notation $[.]_i^{\xi_i}(k_i)$ signifies that the parameter is time-independent but its value changes with the attempt number of the flow-line. However, depending on the requirements or context, any sample parameter can be represented with different forms of notations in various places of this paper.

All the important parameters used in the system model and mathematical formulae are listed in Table-I. However, in this paper, the words like sensor sample, status update, and packet are used interchangeably.

## II. SYSTEM MODEL

In this work, a symmetric IWSAN is considered which consists of $M$ distributed sensors, its corresponding $M$ distributed actuators, and a centralized processor or controller, connecting the sensor-actuator pair through a wireless mesh network, as shown in Fig. 2. Each sensor and its related actuator are indexed by notation $i$, for $i = 1,2,...,M$.

Sensors are monitoring an underlying stochastic process of its interest and updating the controller about its present status. These sensors share a single hop, unreliable channel on a dynamic TDMA basis for transmitting their sensed samples to the controller. Each TDMA slot is denoted by $t$ where $t = 1,2,3...T$. $T$ is the size of the finite (discrete) time horizon. The duration of each slot $t$ lasts from time $t-1$ to $t$. However, this channel has $ON$ probability $p_i = p \in [0,1]$. During this time the channel transmits packets reliably. At any time slot $t$, channel state is either $ON$ or $OFF$ as follows,

$$ch = 1\ (Channel\ ON), \qquad w.p. = p$$

$$= 0\ (Channel\ OFF),\ \ w.p. = (1-p). \qquad (1)$$

At the beginning of each slot $t$, sensors that collect new samples, compete for transmitting their packets to the controller using that common channel. The controller runs a predefined scheduling policy $\pi$ and selects one of them for further processing.

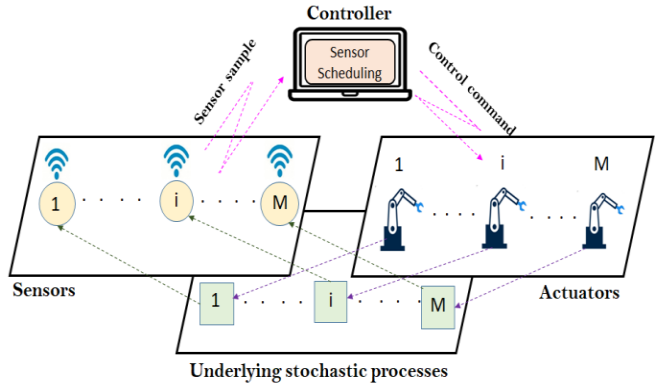

Fig. 2. Symmetric WSAN for closed loop control in ICPSs.

TABLE- I

List of Notations

| Notations | Descriptions |
|---|---|
| $M$ | Number of sensor-actuator pairs in symmetric IWSAN. |
| $i$ for $i = 1,2, \dots M$ | Index for sensor and its corresponding actuator. |
| $T$ | Finite time horizon. |
| $t$ for $t = 1,2, \dots T$ | Index for TDMA slots. |
| $p_i = p \in [0,1]$ | Probability for channel reliability. |
| $k_i$ | Index number of $k^{th}$ sample packet from active sensor $i$. |
| $h_{t,i}(k_i)$ | Age of Information of the sample packet $k_i$ from sensor $i$ at the beginning of slot $t$. |
| $c_i(k_i - 1)$ | Action execution time based on $(k_i - 1)^{th}$ sample from sensor $i$. (Useful age.) |
| $L_{t,i}(k_i)$ | Latency incurred up to time slot $i$ before the successful processing of $k_i^{th}$ sample. |
| $P_i(k_i)$ | Processing time of the controller for the $k_i^{th}$ sample. (In this paper it always takes the value 1). |
| $L_{t,i}(k_i) + P_i(k_i)$ | Stale age. |
| $d^t$ | Active sensor sample scheduled for transmission at any slot $t$. |
| $drop^t$ | Critical sample dropped at the end of the slot $t$. |
| $\xi_i$ | Attempt number of $i^{th}$ flow-line. |
| $RS_i^{\xi_i}(k_i)$ | Reset time for the hibernating flow-line $i$ after $k_i^{th}$ sample from sensor $i$ is dropped at attempt $\xi_i$. |
| $a_i^{\xi_i}(k_i)$ | First arrival epoch of sample $k_i$ during $\xi_i^{th}$ attempt of flow-line $i$. |
| $D_i^{\xi_i}(k_i)$ | Deadline associated with $k_i^{th}$ sample from sensor $i$ at attempt $\xi_i$. |
| $\vec{I}$ | Initial condition vector. |
| $\alpha_i = \alpha > 0 \ \forall \ i \in M$ | Real valued weight assigned to any sensor $i$ in the WSAN. |
| $AoI_{t,i}$ | Age incurred by the sensor $i$ during one slot $t$. |
| $AoI_i$ | Total Age incurred by the sensor $i$ during the total finite time horizon T. |
| $\overline{AoI}$ | Average age of information in the network for M sensor-actuator pairs and finite time horizon T. |
| $AL$ | Average latency of sensor samples in the network for M sensor-actuator pairs and time horizon T. |
| $JT$ | RMS jitter of the actuators in the network for M sensor-actuator pairs and finite time horizon T. |
| $U_{t,i}(k_i)$ | Utility of the information content in a sample $k_i$ from a sensor $i$ at any slot $t$. |
| $F_{t,i}(k_i)$ | Freshness of the information content in a sample $k_i$ from a sensor $i$ at any slot $t$. |
| $LX_{t,i}^{\xi_i}(k_i)$ | Laxity of sample $k_i$ from a sensor $i$ at any slot $t$ in its $\xi_i^{th}$ attempt. |
| $X_{t,i}(k_i)$ | Influence of laxity for any sample $k_i$ on its utilization at slot $t.$. |
| $\mathbb{P}_{t,i}(k_i)$ | Penalty on $k_i$ sample from sensor $i$ at any time slot $t$. |
| $Lt_{t,i}^{\xi_i}(k_i)$ | Lateness (tardiness) of sample $k_i$ from a sensor $i$ at any slot $t$ in its $\xi_i^{th}$ attempt. |
| $\psi\left(Lt_{t,i}^{\xi_i}(k_i)\right)$ | Effect of lateness for sample $k_i$ from a sensor $i$ at any slot $t$ in its $\xi_i^{th}$ attempt. |
| $EU_{t,i}(k_i)$ | Effective utility (utilization) of information present in a sample $k_i$ from a sensor $i$ at any slot $t$. |
| $Pri_{t,i}(k_i)$ | Priority of the active sensor $i$. |
| $EXWSUoI$ | Expected weighted sum utility of information of the network. |
| $S_t$ | Set of active sensors at any slot $t$. |
| $S_t'$ | Modified version of set $S_t$ arranged in decreasing order of priority of the constituent active sensors. |
| $R^{(t)}$ | Number of elements present in $S_t$ (or, $S_t'$). |
| $\{x_a\}_t$ for $a = 1 \dots R^{(t)}$ | Elements (active sensors) in $S_t$. |
| $\{e_{a'}\}_t$ for $a' = 1 \dots R^{(t)}$ | Elements in $S_t'$. |
| $Pri_{t,a}$ | Priority of the active sensor $x_a$ in $S_t$. |
| $Pri_{t,a'}$ | Priority of the active sensor $e_{a'}$ in $S_t'$. |
| $U_{t,a}$ | Utility of the active sensor $x_a$ in $S_t$. |
| $U_{t,a'}$ | Utility of the active sensor $e_{a'}$ in $S_t'$. |
| $N_{t+1} = \{y_n\}_{t+1}$ for $n = 1 \dots r^{(t+1)}$ | Set of newly active sensors at the beginning of slot $t$. |
| $RN_{t+1} = \{\chi_\eta\}_{t+1}$ for $\eta = 1 \dots \varepsilon^{(t+1)}$ | The set of regenerated sensors at slot $t + 1$, arranged in the decreasing order of their priority. |
| $\{y_n\}_t$ for $n = 1 \dots r^{(t)}$ | Elements in $N_t$. |
| $V_t$ | Sum of utilization of all the active sensor samples present in $S_t$ (or, $S_t'$). |

After the successful processing of a sensor sample, the controller instantaneously and reliably sends an acknowledgment (ACK) to that sensor.

AoI increases linearly with the time elapsed since the arrival epoch of the most recent successfully processed sample, as shown in Fig 3. Only after the successful processing of one sample at any time slot $t$, its AoI is updated to 1 at the end of that slot. Otherwise, at the end of the slot, AoI of the unprocessed active sensor sample increases by one slot time and it senses the current status of the system again in the next slot by replacing the old sample. So, **our definition of 'age of information' captures the number of time slots passed after the sensing of last successful processed status update containing the information about the system state.** If $d^t$ be the active sensor whose sample packet has been processed successfully at slot $t$, at the end of that slot, AoI evolves as follows,

$$h_{t+1,i}(k_i) = h_{t,i}(k_i) + 1 \; if \, i \neq d^t$$

$$h_{t+1,i}(k_i + 1) = 1 \; if \, i = d^t. \tag{2}$$

Here, $h_{t,i}(k_i) \geq 0$ is the age of the information content present in the sample from sensor $i$ at the beginning of slot $t$. $k_i$ is the index number of that sample packet. If $k_i^{th}$ packet from any active sensor $i$ is processed successfully, packet index changes from $k_i$ to $k_i + 1$.

Upon receipt of the sample from sensor $i$, the controller analyzes the information content present in that sample and generates an appropriate control command. Then, within the same slot $t$, the command packets are reliably sent to the corresponding actuator $i$ for performing necessary control actions. So, the controller takes one slot time in total for sensor scheduling, control decision making, and transmission of control command to the actuator [38].

Actuator takes time equivalent to $c_i(k_i) \geq 0$ number of slots for the execution of the assigned actuation task. It is quite obvious that only after the completion of the previously assigned task, the present state of the system changes significantly and the sensor $i$ needs to sense this change. So, after the successful processing of $k_i^{th}$ sample from sensor $i$, it waits for $(c_i(k_i) + 1)$ slots time, before sensing the next sample $(k_i + 1)$ and participating in scheduling again. This $c_i(k_i)$ amount of slot time for action execution is termed as the '*setup time*' of the $i^{th}$ flow-line for its $(k_i + 1)^{th}$ sample. Until the age value reaches $(c_i(k_i) + 1)$, sensor $i$ is said to be '*inactive*' or in '*sleep*' mode. It neither picks any sample nor takes part in the scheduling. Significance of this mode lies in its ability to protect the network from unnecessary congestion and exploitation of network resources (bandwidth, power, etc.) by avoiding redundant status updates. Moreover, this simplifies the scheduling decision by excluding inactive sensors from the competition.

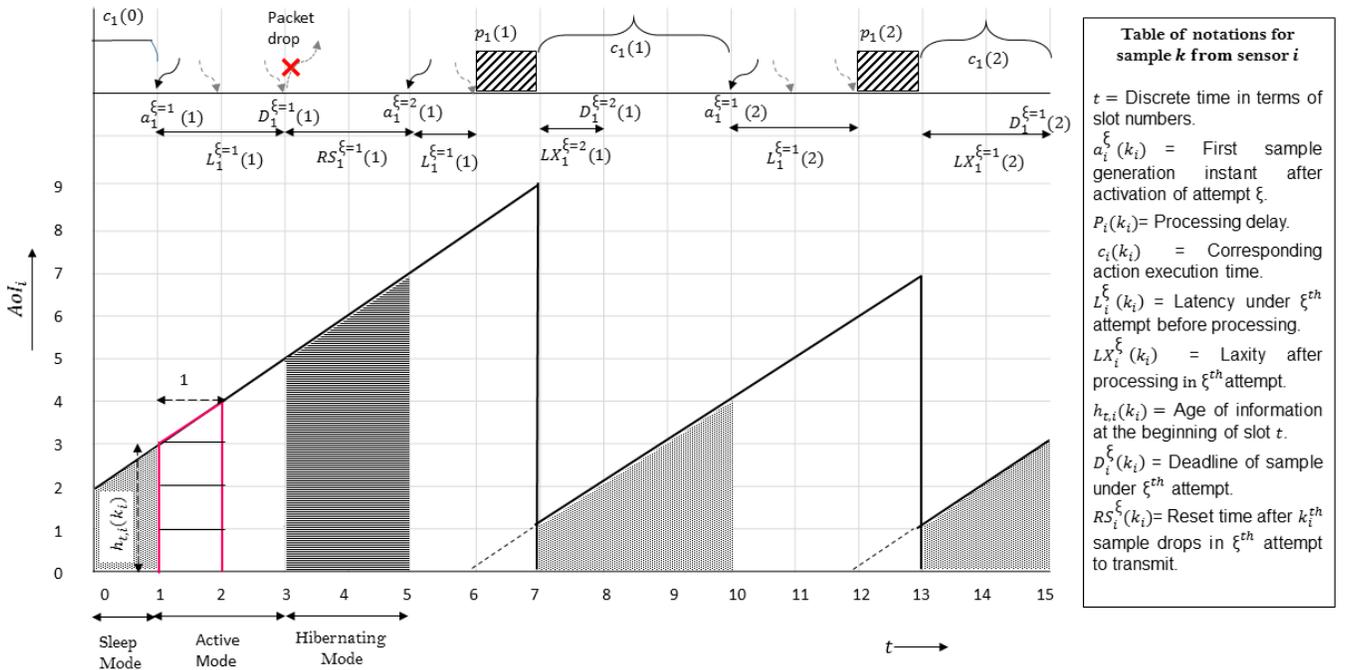

Fig. 3. Age evolution for sensor $i$.

Only when the $k_i^{th}$ sample from sensor $i$ is having age $h_{t,i}(k_i)$ greater than its '*setup time*', it is called the '*active*' sensor. It senses samples periodically in each slot with the replacement of the older unprocessed sample and participates in scheduling until its newly sensed sample is being processed successfully. But, this sensor cannot attempt to get service for an indefinite number of times. Each sensor has a deadline $D_i(k_i)$ associated with its $k_i^{th}$ sample. This deadline is always greater than or equals to the '*setup time*' of $k_i^{th}$ sample as only after this much amount of age, a new sample $k_i$ is being sensed. So, with modification, we can say that a sensor is considered to be active only when its age is greater than the '*setup time*' but not exceeding the deadline.

When the age of information $h_{t,i}(k_i)$ of sample $k_i$ crosses its deadline, the packet from sensor $i$ is dropped and as a result, the $i^{th}$ flow-line goes out of service causing production loss or even more fatal consequences. This flow-line, then, requires some additional servicing and maintenance to become functional again in order to sense the sample and participate in scheduling. This restoration process takes some arbitrary amount of time of $RS_i(k_i)$ slots, known as the '*reset time*' *of the* $i^{th}$ flow-line after the drop of its $k_i^{th}$ sample. This temporary blackout period of the flow-line is called the '*hibernation*' mode and for very obvious reason, this is not at all a desired situation to occur. However, after a flow-line enters '*hibernation*', its underlying incomplete task is controlled (or, eventually shut down, if applicable) by some default or preset control commands.

After reset, the sensor from flow-line $i$ becomes active again with age $(D_i(k_i) + RS_i(k_i) + 1)$ slots and reattempts to transmit a newly sensed sample $k_i$ having a new deadline. In other words, if a sensor fails to transmit its packet $k_i$ at its $\xi_i^{th}$ attempt and the packet is dropped, then, after reset, the attempt number for the sensor to transmit its fresh $k_i$ sample changes from $\xi_i$ to $\xi_i + 1$. The new deadline of this sample is assigned as, $D_i^{\xi_i+1}(k_i) \geq D_i^{\xi_i}(k_i) + RS_i^{\xi_i}(k_i) + 1$, where $\xi_i = 1,2,3, ..., RS_i^{\xi_i}(k_i)$ is the '*reset time*' after the failure of $\xi_i^{th}$ attempt to transmit the $k_i^{th}$ sample from sensor $i$, and $D_i^{\xi_i}(k_i)$ is the deadline of $k_i^{th}$ sample at its $\xi_i^{th}$ attempt to transmit. Attempt number $\xi_i = 0$ indicates no active packet is available in the sensor $i$. Then, $D_i^0(k_i) = 0$ and $RS_i^0(k_i) = c_i(k_i - 1) + 1$.

Finally, it can be concluded that the total age $h_{t,i}(k_i)$ of any sample $k_i$ from a sensor $i$ is the sum of action execution time $c_i(k_i - 1)$ of the previously processed sample $k_i - 1$ from the same sensor, latency or waiting time $L_{t,i}(k_i) \geq 0$ of the current sample $k_i$ before its successful processing and its processing delay $P_i(k_i) \geq 0$ [34]. Latency and processing delay together result in the end to end communication delay. During initial $c_i(k_i - 1)$ amount of age, some useful actuation task takes place and the sensor lies in sleep mode. So $c_i(k_i - 1)$ is considered to be '*useful*' content in $h_{t,i}(k_i)$ and we can not control it externally. Remaining $\Delta_i = L_{t,i}(k_i) + P_i(k_i)$ amount of time is called '*stale age*' as it is unnecessarily degrading the system performance by prolonging end to end communication delay as well as 'age'. As discussed earlier in this system model processing delay is always constant one slot time.

$$h_{t,i}(k_i) = \underbrace{c_i(k_i - 1)}_{Useful\ age} + \underbrace{L_{t,i}(k_i) + P_i(k_i)}_{Stale\ age}. \tag{3}$$

At any time slot $t$, depending on the age $h_{t,i}(k_i)$ and the attempt number $\xi_i$ of the flow-line, any sensor $i$ lies in one of the three following modes,

*Inactive:* 
$$1 \leq h_{t,i}(k_i) < c_i(k_i - 1) + 1 = RS_i^0(k_i) \tag{4}$$

*Active:* 
$$D_i^{\xi_i - 1}(k_i) + RS_i^{\xi_i - 1}(k_i) \leq h_{t,i}(k_i) < D_i^{\xi_i}(k_i) \tag{5}$$

*Hibernation:* 
$$D_i^{\xi_i}(k_i) \leq h_{t,i}(k_i) < D_i^{\xi_i}(k_i) + RS_i^{\xi_i}(k_i). \tag{6}$$

## III.   MATHEMATICAL FORMULATIONS

Evolution of the age of sensor $i$ for this proposed system model is illustrated in Fig. 3. The total age $AoI_i$ is obtained from the area covered by the age curve of sensor $i$, can be obtained as follows,

$$AoI_i = \sum_{t=1}^{T} AoI_{t,i} = \frac{1}{2} + \sum_{t=1}^{T} h_{t,i}(k_i). \tag{7}$$

The average age of the system will be,

$$\overline{AoI} = \frac{1}{TM}\sum_{t=1}^{T}\sum_{i=1}^{M} AoI_{t,i}$$

$$= \frac{1}{TM}\sum_{t=1}^{T}\sum_{i=1}^{M}\left\{\frac{1}{2} + h_{t,i}(k_i)\right\}$$

$$= \frac{1}{2} + \frac{1}{TM}\sum_{t=1}^{T}\sum_{i=1}^{M}\{h_{t,i}(k_i)\}. \tag{8}$$

As the constant terms does not affect the optimization problem, therefore, neglecting the constant part slightly modified expression for (8) will be,

$$\overline{AoI} \approx \frac{1}{TM}\sum_{t=1}^{T}\sum_{i=1}^{M}\{h_{t,i}(k_i)\}$$

$$\approx \frac{1}{TM}\sum_{t=1}^{T}\sum_{i=1}^{M}\{c_i(k_i - 1) + L_{t,i}(k_i) + P_i(k_i)\}. \tag{9}$$

Next, neglecting the constant parts $c_i(k_i - 1)$ and $P_i(k_i)$, average latency (AL) is obtained as follows,

$$AL = \frac{1}{TM}\sum_{t=1}^{T}\sum_{i=1}^{M}\{L_{t,i}(k_i)\}. \tag{10}$$

RMS jitter present in the actuation tasks is calculated as,

$$JT = \sqrt{\frac{1}{TM}\sum_{t=1}^{T}\sum_{i=1}^{M}\{L_{t,i}(k_i)\}^2}. \tag{11}$$

From the existing literature, it is known that AoI measures the freshness of information. A smaller value of age ensures a higher freshness of information. However, from eqn. (3) we can say that removal of the stale part from the age, in turn, reduces the total age of information. From our previous work [34] it can be concluded that for ICPS (as described in system model in sec II) minimizing average latency (or staleness) that is inherently involved in age metric is more effective than minimizing the total age value in case of handling both cyber and physical parts simultaneously as a closed-loop networked control system. Mitigating staleness, in turn, assures the freshness of information along with smooth execution of the physical processes by minimizing the jitter. Moreover, from the above system model it is known that whenever a packet misses its deadline, the corresponding flow-line goes out of service temporarily (hibernation mode) which is not at all expected. So basically we have two objectives to be taken care of conjointly:

1. To minimize the average latency of a sensor sample.
2. To minimize the number of samples that misses their hard deadline.

These two objectives can be met simultaneously by satisfying only one equivalent objective that is: **maximizing the average usefulness/utility of the information contained in an active sensor sample.**

Let, the utility of information content in a sample $k_i$ from a sensor $i$ be $U_{t,i}(k_i)$ at any slot $t$. $U_{t,i}(k_i)$ is a function of two terms, i.e. freshness of information $F_{t,i}(k_i)$ and laxity/slack time $LX_{t,i}^{\xi_i}(k_i)$ of sample $k_i$ with attempt number $\xi_i$ at slot $t$.

$$U_{t,i}(k_i) = f\left(F_{t,i}(k_i), X_{t,i}(k_i)\right)$$

$$where, X_{t,i}(k_i) = \mathfrak{I}\left(LX_{t,i}^{\xi_i}(k_i)\right). \tag{12}$$

Freshness $F_{t,i}(k_i)$ is a positive, bounded, real-valued, bivariate monotonically decreasing function with increasing latency $L_{t,i}(k_i)$ and attempt number $\xi_i$ of the sensor sample $k_i$ that contains the information about the system state.

$$F_{t,i}(k_i) = g\left(L_{t,i}(k_i), \xi_i\right)$$

$$s.t: F_{t,i}(k_i) < F_{t,j}(k_j) \text{ when, } L_{t,i}(k_i) > L_{t,j}(k_j) \text{ or } \xi_i > \xi_j. \tag{13}$$

As shown in Fig 3, at any slot $t$, if the active sensor $i$ is trying to transmit its unprocessed sample at its $\xi_i^{th}$ attempt and $a_i^{\xi_i}(k_i) \leq t \leq D_i^{\xi_i}(k_i)$, then the latency of the active sample $k_i$ for this attempt will be, $L_{t,i}^{\xi_i}(k_i) = \left( t - a_i^{\xi_i}(k_i) \right)$. The total latency of the sample is considered to be the time passed since the first arrival epoch of the sample $k_i$ before the first service attempt starts.

$$L_{t,i}(k_i) = t - a_i^1(k_i) = L_{t,i}^{\xi_i}(k_i) + D_i^{\xi_i - 1}(k_i). \tag{14}$$

When $L_{t,i}^{\xi_i}(k_i) = 0$ and $\xi_i = 1$, freshness $F_{t,i}(k_i)$ reaches its maximum limit 1. $U_{t,i}(k_i)$ is directly proportional to any real power of $F_{t,i}(k_i)$. So, from (13), $U_{t,i}(k_i)$ is also a bivariate monotonically decreasing function of $(L_{t,i}(k_i), \xi_i)$. In this paper, $F_{t,i}(k_i)$ is considered as $\frac{1}{(\xi_i - 1)D_{max} + L_{t,i}^{\xi_i}(k_i) + 1}$. Here $D_{max}$ is the maximum possible value of the deadline of any active sensor sample.

Laxity or slack time of a sensor sample is defined as the remaining time up to its deadline after the successful processing of the sample, starting from the current instant. From Fig 3, if $k_i^{th}$ sample be successfully processed during its $\xi_i^{th}$ attempt at any slot $t$ and $a_i^{\xi_i}(k_i) \leq t \leq D_i^{\xi_i}(k_i)$, the laxity of $k_i^{th}$ sample at slot $t$ will be, $LX_{t,i}^{\xi_i}(k_i) = (D_i^{\xi_i}(k_i) - t) = (D_i^{\xi_i}(k_i) - L_{t,i}^{\xi_i}(k_i) - 1)$ slots time (as processing takes only one slot time in this case). So, when $L_{t,i}^{\xi_i}(k_i) = 0$, $LX_{t,i}^{\xi_i}(k_i) = (D_i^{\xi_i}(k_i) - 1)$ and when $L_{t,i}^{\xi_i}(k_i) = (D_i^{\xi_i}(k_i) - 1)$, $LX_{t,i}^{\xi_i}(k_i) = 0$. So, we get $0 \leq L_{t,i}^{\xi_i}(k_i) < D_i^{\xi_i}(k_i)$ and $0 \leq LX_{t,i}^{\xi_i}(k_i) < D_i^{\xi_i}(k_i)$. That implies if any sample is having latency $L_{t,i}^{\xi_i}(k_i) = \left( D_i^{\xi_i}(k_i) - 1 \right)$ and still, it is not scheduled to be processed then it will definitely be dropped and the corresponding flow-line will enter the hibernation mode. This type of sample with zero laxity is called '*critical Sample*'. An active sensor having the critical sample is called '*critical sensor*' and an active sensor with non-critical sample is called as '*non-critical sensor*'.

Now, as described in the system model, the processor can serve at most one sensor sample in each slot. For this constraint, if more than one critical samples are present at any slot, irrespective of the undergoing scheduling algorithm, at most one critical sample can be saved from dropping at that particular slot. All other critical samples will be dropped and their corresponding flow-line will go out of service. To avoid such a situation, if more than one critical sample is present at a particular slot $t$, only one critical sample is considered to have the hard deadline and other critical samples are being graced by delaying their deadlines by one more slot. The same phenomena will take place when any critical sample appears but the channel is $OFF$ at that particular slot. Irrespective of the scheduling policy in use, no sample can be processed at that slot and, as a result, all the critical samples will certainly be dropped. To avoid such a situation whenever the transmission channel is $OFF$, critical samples are considered to have soft deadlines and their deadlines are being graced by one more slot to offer them a chance to retransmit. In this way, unnecessary packet drops due to conflict among multiple critical samples or due to bad channel conditions can be circumvented. This is called the 'Conflict Avoidance' (CA) mechanism. But, due to this mechanism, a critical sensor with soft deadline faces tardiness which in turn incurs some penalty value $\mathbb{P}_{t,i}(k_i)$. $\mathbb{P}_{t,i}(k_i)$ is being added to the utilization of the sensor sample which is being benefitted by CA mechanism. So, its effective utilization ($EU_{t,i}(k_i)$) is $EU_{t,i}(k_i) = (U_{t,i}(k_i) - \mathbb{P}_{t,i}(k_i))$.

This penalty function $\mathbb{P}_{t,i}(k_i)$ is nothing but the effect of tardiness (lateness) on the current utilization of the sensor sample. For an active sensor sample $k_i$ from sensor $i$, the penalty at any slot $t$, can be represented as follows,

$$\mathbb{P}_{t,i}(k_i) = U_{t,i}(k_i) * \psi \left( Lt_{t,i}^{\xi_i}(k_i) \right) * \mathbb{I}_+ \left( Lt_{t,i}^{\xi_i}(k_i) \right). \tag{15}$$

Here, $Lt_{t,i}^{\xi_i}(k_i)$ is the lateness of the sample $k_i$ for its initial deadline $D_i^{\xi_i}(k_i)$ (without gracing) and $\xi_i$ transmission attempt for $i^{th}$ flow-line. $\mathbb{I}_+(.)$ is an indicator function that takes value 1 for $Lt_{t,i}^{\xi_i}(k_i) > 0$ else $\mathbb{I}_+ \left( Lt_{t,i}^{\xi_i}(k_i) \right) = 0$. Penalty decreases with increasing tardiness and/or decreasing utilization. This is because the increasing tardiness makes the sensor update outdated or stale whereas higher utilization signifies that the sample packet is carrying some urgent or very useful message whose loss may cause some fatal consequences. Therefore, during CA, our aim is to choose a packet for service having higher effective utilization $EU_{t,i}$ at that time. In Fig 4 we have plotted the penalty function considering $\psi \left( Lt_{t,i}^{\xi_i}(k_i) \right)$ as $e^{-\frac{10*ln\left( T - (\xi_i - 1)D_{max} - D_i^{\xi_i} - Lt_{t,i}^{\xi_i} \right)}{T}}$. This $\psi \left( Lt_{t,i}^{\xi_i}(k_i) \right)$ belongs to the family of those functions whose product with $U_{t,i}(k_i)$ follow the nature of our desired penalty property, as discussed above.

Whenever a critical sample drops, its utility becomes 0. Any positive laxity of a sample does not affect its utility. So it can be expressed as follows,

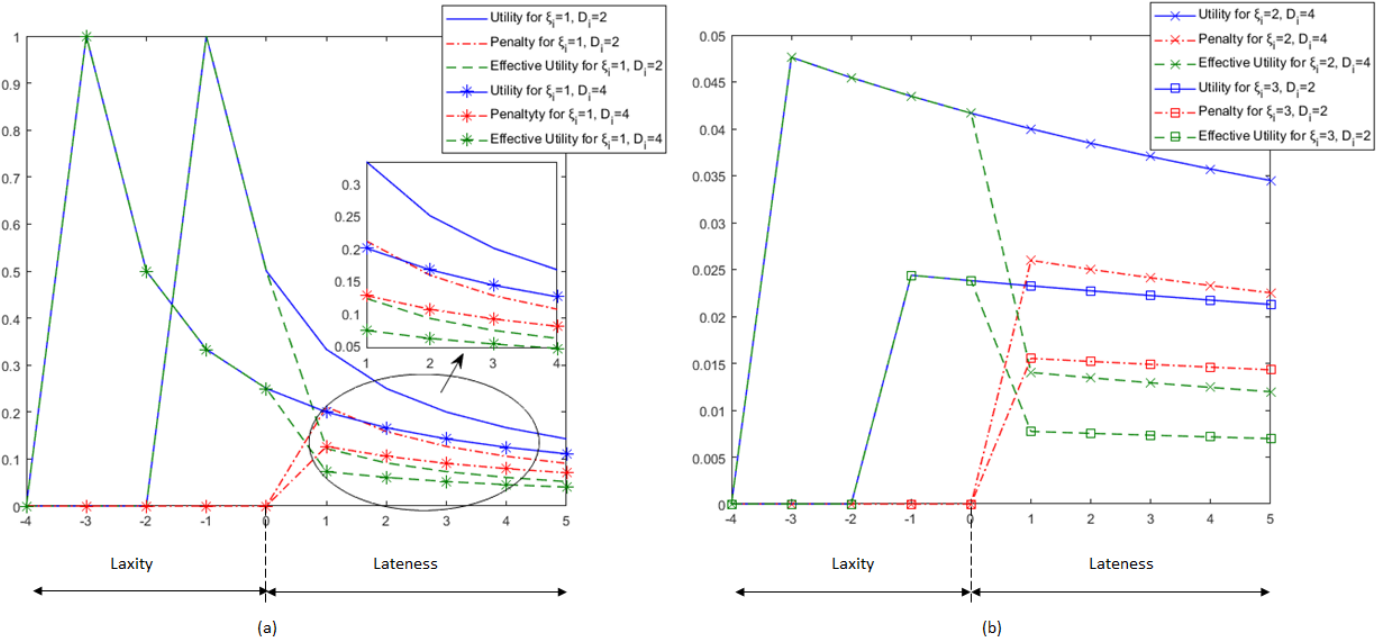

Fig. 4. Comparisons of utility, penalty, and effective utility for T = 100, $D_{max} = 20$, and (a) same $\xi_i$ but different values of $D_i$ (b) different values of $\xi_i$ and $D_i$.

$$X_{t,i}(k_i) = \mathfrak{F}\left(LX_{t,i}^{\xi_i}(k_i)\right)$$

$$X_{t,i}(k_i) = \begin{cases} 0 & for\ LX_{t,i}^{\xi_i}(k_i) < 0 \\ 1 & otherwise. \end{cases} \quad (16)$$

$U_{t,i}(k_i)$ is directly proportional to any real power of $X_{t,i}(k_i)$. So, the utility can be expressed as,

$$U_{t,i}(k_i) = k_F k_X F_{t,i}^{\beta}(k_i) X_{t,i}^{\gamma}(k_i) = k F_{t,i}^{\beta}(k_i) X_{t,i}^{\gamma}(k_i). \quad (17)$$

Here, $k_F, k_X$ are proportional constants for $F_{t,i}(k_i)$ and $X_{t,i}(k_i)$, respectively and $k = k_F k_X$. $\beta, \gamma$ are real indices.

Next, we introduce a special notion of '$Priority$' ($Pri_{t,i}(k_i)$) of a sensor $i$ at any time slot $t$, based on the utility ($U_{t,i}(k_i)$) and laxity ($LX_{t,i}(k_i)$) it's sample $k_i$ offers. It is assigned as follows:

$$Pri_{t,i}(k_i) = \begin{cases} \dfrac{1}{U_{t,i}(k_i)\ X_{t+1,i}(k_i)} > 0 & for\ X_{t+1,i}(k_i) = 1 \\ \infty & for\ X_{t+1,i}(k_i) = 0 \end{cases}$$

$$where,\ X_{t+1,i}(k_i) = \mathfrak{F}\left(LX_{t+1,i}^{\xi_i}(k_i)\right)\ and\ U_{t,i}(k_i) > 0. \quad (18)$$

If a sample $i$ is critical at $t$, then $LX_{t,i}^{\xi_i}(k_i) = 0$ and $LX_{t+1,i}^{\xi_i}(k_i) < 0$. So, $X_{t+1,i}(k_i) = 0$ and $Pri_{t,i}(k_i) = \infty$. On the other hand, if a sample $i$ is non-critical at $t$, then $LX_{t,i}^{\xi_i}(k_i) > 0$ and $LX_{t+1,i}^{\xi_i}(k_i) \geq 0$. So, $X_{t,i}(k_i) = X_{t+1,i}(k_i) = 1$, $Pri_{t,i}(k_i) = \frac{1}{U_{t,i}(k_i)} = \frac{1}{k F_{t,i}^{\beta}(k_i) X_{t,i}^{\gamma}(k_i)} = \frac{1}{k F_{t,i}^{\beta}(k_i)} = \frac{1}{k\{g(L_{t,i}(k_i),\xi_i)\}^{\beta}}$.

That means, when a sample is non-critical its priority monotonically increases with decreasing utilization (or freshness). But, whenever a packet is critical, its priority to be chosen for processing is becoming the highest always.

The average utility of the information content in the system associated with M sensors is,

$$U = \frac{1}{TM} \sum_{t=1}^{T} \sum_{i=1}^{M} U_{t,i}(k_i)$$

$$= \frac{k}{TM} \sum_{t=1}^{T} \sum_{i=1}^{M} F_{t,i}^{\beta}(k_i) X_{t,i}^{\gamma}(k_i)$$

$$= \frac{k}{TM} \sum_{t=1}^{T} \sum_{i=1}^{M} \{g(L_{t,i}(k_i), \xi_i)\}^{\beta} \{\mathcal{I}\left(LX_{t,i}^{\xi_i}(k_i)\right)\}^{\gamma}. \qquad (19)$$

Now, if any sensor $i \in M$, be in sleep or hibernating mode, it refrains from participating in scheduling. Therefore, applying scheduling policy only on the set of active sensors is sufficient to maximize the utilization. Let, $S_t$ be the set of all the active sensors at the beginning of any slot t. In order to maximize the utility of information, the expected weighted sum utility of information (EXWSUoI) is chosen as the objective function with finite horizon $T$, as follows:

$$EXWSUoI = \frac{k}{TM} E\left[\sum_{t=1}^{T} \sum_{i \in S_t} \{\alpha_i U_{t,i}(k_i) | \vec{I}\}\right]. \qquad (20)$$

Here, $\vec{I} = [\overrightarrow{h_1(1)}, \overrightarrow{c(0)}, \overrightarrow{D(1)}]$ is the initial condition vector that represents the initial values of $h_{t,i}(k_i)$, $c_i(k_i - 1)$ and $D_i(k_i)$ respectively for $k_i = 1$. $\alpha_i = \alpha > 0 \ \forall \ i \in M$ is the real-valued weight assigned to any sensor $i$ in the WSAN. Henceforth, initial condition and packet index will be omitted for notation simplicity.

**Finally, this paper focuses on finding suitable offline, work conserving, energy efficient, dynamic priority-based, deadline-aware scheduling policy $\pi$ from the set of all the admissible policies $\Pi$ to maximize the objective function.**

**Objective Function:**
$$O_T^{\pi^*} = \max_{\pi \in \Pi} E[O_T^{\pi}],$$

$$where, O_T^{\pi} = \frac{k\alpha}{TM} \left[\sum_{t=1}^{T} \sum_{i \in S_t} U_{t,i}\right].$$

## IV. OPTIMALITY OF DEADLINE-AWARE HIGHEST LATENCY FIRST ALGORITHM

In this section, first, a greedy policy **Deadline-aware Highest Latency First (HLF-D)** is discussed. This algorithm was introduced in our previous work [38] to maximize the freshness of information in the absence of the flow-line regenerations. In this section, the optimality of HLF-D is analyzed and proved that this algorithm also provides the global optimal result for our finite-horizon scheduling problem with flow-line regeneration by maximizing the average utility of information while satisfying the objective function (described in sec III).

In [38] HLF-D algorithm is defined as follows:

**Definition:** *(i) At any time slot, if no critical sample is present in the system, HLF-D, a greedy scheduling policy, schedules the transmission of an unprocessed sample from the active sensor with the highest latency. (ii). The presence of a critical sample ensures that none but the critical sample will get service. (iii) [Tie-Breaking Condition] All ties are being broken arbitrarily.*

Next, in this section, a theorem followed by its supporting lemma is discussed that proceeds towards the optimality proof of the algorithms.

**Lemma-I:** *Any active, non-critical sensor sample having higher utility at a particular slot, offers higher utility value in the next slot, as well, if not served.*

**Proof:** At any slot $t$, let, the sample from an active, non-critical sensor $i$ has higher utility value $U_{t,i}$ than that of any other sample from active sensor $j$. We have to prove that, $U_{t+1,i} \geq U_{t+1,j}$.

Now, $U_{t,i} = k\{g(L_{t,i}, \xi_i)\}^{\beta} X_{t,i}^{\gamma}$. $L_{t+1,i} = L_{t,i} + 1 \ if \ i \neq d^t$ and $g(L_{t,i} + 1, \xi_i) < g(L_{t,i}, \xi_i)$. $X_{t,i} = 1$ when $i$ is active and $X_{t+1,i} = 1$ too if $i$ is non-critical at $t$. So, $U_{t+1,i} = k\{g(L_{t+1,i}, \xi_i)\}^{\beta} X_{t+1,i}^{\gamma} = k\{g(L_{t+1,i}, \xi_i)\}^{\beta} < k\{g(L_{t,i}, \xi_i)\}^{\beta} = U_{t,i}$. If sample $j$ is critical at slot $t$, then, in the next slot $t + 1$, its utilization $U_{t+1,j}$ will become 0 whereas $U_{t+1,i} \geq 0$ always. However, if sample $j$ is non-critical at slot $t$,

$$U_{t,i} \geq U_{t,j} \text{ (Statement of Lemma-I)}$$

$$or, \quad \{g(L_{t,i}, \xi_i)\}^\beta \geq \{g(L_{t,j}, \xi_j)\}^\beta$$

$$or, \quad \{g(L_{t,i} + 1, \xi_i)\}^\beta \geq \{g(L_{t,j} + 1, \xi_j)\}^\beta$$

$$or, \quad U_{t+1,i} \geq U_{t+1,j}. \qquad\qquad (\because \text{Proved})$$

***Theorem-I:*** *For any symmetric IWSAN network with an unreliable time-shared channel from the sensor to the processor, among the class of admissible policies, HLF-D attains the maximum expected weighted sum utility of information ($O_T^{\pi^*}$) for active sensor samples with deadlines while the flow-line regeneration takes place, if any packet drops.*

***Proof:*** The system model is described in sec II. The *ON-OFF* condition of the time-shared channel is time-varying and at any slot $t$, it is neither dependent on participating sensors nor on the designated scheduling policy. In order to prove the optimality of HLF-D scheduling algorithm, the value of the objective function ($E[O_T^\pi]$), obtained by HLF-D and any other admissible policy $\pi \in \Pi$ are compared.

Let $V_t = \sum_{i \in S_t} U_{t,i}$ be the sum of the utilities of all the active sensor samples in the active sensor set $S_t$ at slot $t$. If it can be proved that $V_t^{HLF-D} \geq V_t^\pi \ \forall t \in T$, then it is sufficient to state that $O_T^{\pi^*} = E[O_T^{HLF-D}] \geq E[O_T^\pi] \ \forall \pi \in \Pi$. That means HLF-D globally maximizes the expected usefulness of the information while the flow-line regeneration is taken into consideration.

Now, let us denote the elements in $S_t$ as $x_a$ for $a$ in the range [1, M]. So, $S_t = \{x_a\}_t$, $a = 1 \ldots R^{(t)}$ be the set of total $R^{(t)}$ number of active sensors at slot $t$. $U_{t,a}$ and $Pri_{t,a}$ are the utilization and priority, respectively, of any sensor sample $i$ belongs to $S_t$. Let, $S_t' = \{e_{a'}\}_t$, $a' = 1 \ldots R^{(t)}$ is the modified set $S_t$ arranged in decreasing order of priority of the constituents. Utility and priority of any element $a' \in S_t'$ are denoted as $U_{t,a'}$ and $Pri_{t,a'}$, respectively.

Let, the sample $d^t$ from the reshuffled active sensor set $S_t'$ is processed successfully at slot $t$. So at the end of slot $t$, latency will be $L_{t+1,a'} = L_{t,a'} + 1$ and laxity will be $LX_{t+1,a'}^{\xi_{i'}} = LX_{t,a'}^{\xi_{i'}} - 1$ if $a' \neq d^t$. If any critical sample is present in slot $t$ and it is not scheduled to be processed, then this sample will be dropped at the end of this slot (denoted by $drop^t$). Sensor, from which critical sample is dropped, join the active set again after the reset time with a fresh sensed sample. So, in the next slot $t + 1$, active set $S_{t+1} = [S_t' \backslash \{d^t, drop^t\}] \cup N_{t+1} \cup RN_{t+1}$ where, $N_{t+1} = \{y_n\}_{t+1} \ for \ n = 1 \ldots r^{(t+1)}$ be the set of $r^{(t+1)}$ new sensors that just enter the active set at the beginning of slot $t + 1$. The latency of any sensor sample $n \in N_{t+1}$ be $L_{t+1,i}^1 = L_{t+1,n}^1 = 0$ and utility $U_{t+1,i} = U_{t+1,n} = 1$. $RN_{t+1} = \{\chi_\eta\}_{t+1} \ for \ \eta = 1 \ldots \ \varepsilon^{(t+1)}$ is the set of regenerated sensors at slot $t + 1$. Elements in regenerated sensor set are arranged in the decreasing order of their priority. Regenerated sensors are those who return to the active set after the reset of their flow-line to reattempt the transmission of their fresh sensor samples once the critical packets are dropped from them at times prior to $t$. As this reset time is arbitrary for any flow-line, multiple hibernating sensors can renew together and join the active set $S_{t+1}$ although they have been dropped at different time slots. They represented collectively as set $RN_{t+1}$. Although, the elements in $RN_{t+1}$ presently having zero latency for their current attempt i.e. $L_{t,i}^{\xi_i} = L_{t,\eta}^{\xi_\eta} = 0$ for $\eta \in RN_{t+1}$, but their attempt number has increased by +1 from the previous one. So, their overall utilization has also reduced accordingly.

We put a 0 for the empty place of $drop^t$ at the beginning of the set $S_{t+1}$. This zero-padding implicates that, according to our system model, the particular flow-line $drop^t$ will enter hibernating mode and its corresponding actuation task will be handled temporarily by some default command from the controller. However, this sensor will stay in the active set as a dummy element with a sample having $utility = 0$ and $laxity < 0$. But, they do not take part in the scheduling. However, regenerated sensors enter the active set $S_{t+1}$ by replacing 0s starting from the last 0's position. Therefore, the number of elements in $S_{t+1}$ becomes $R^{(t+1)} = [max(0, (R^{(t)} - 1)) + r^{(t+1)} + \varepsilon^{(t+1)}] \leq M$, if the channel is *ON* at $t$. If the channel be *OFF* at at $t$, $R^{(t+1)}$ with be same as $R^{(t+1)}$ with $r^{(t+1)}$ and $\varepsilon^{(t+1)}$ number of elements added to it.

Now, if any active sensor with critical sample ($utility > 0$, $laxity = 0$ and $priority = \infty$) pops up in this slot $t + 1$, we place that sensor in $S_{t+1}'$ before those dummy zeros followed by the active sensors with non-critical samples in the decreasing order of their priority.

To prove $V_t^{HLF-D} \geq V_t^\pi \ \forall t \in T$, without the loss of generality, we next try to show that at any slot $t \in T$ for any active sensor $i' \in S_t'^{HLF-D}$, its utility $U_{t,i'}^{HLF-D}$ is higher than that of the $i'^{th}$ sensor sample present in $S_t'^\pi$. So finally our decomposed objective is to prove $U_{t,i'}^{HLF-D} \geq U_{t,i'}^\pi \ \forall i' \in S_t'$ and $t \in T$.

*Assumptions:*

I. For the comparison of the proposed scheduling policy HLF-D with any other policy $\pi \in \Pi$, it is imperative that the number of elements present in the set of newly active sensors $N_{t+1}$, added at the beginning of each slot $t$, remains same irrespective of the underlying scheduling policy.

II. In this proof sensor and sensor sample both the terms can be used interchangeably. This is because the samples are present within the sensor itself.

We are using the **Induction method** for this proof.

***Base case:*** Initial conditions $\vec{I}$ are the same for both the policies. So, the set of active sensors $S_t$ is the same for both HLF-D and $\pi$ at slot $t = 1$. This yields, $U_{1,a'}^{HLF-D} = U_{1,a'}^{\pi} = U_{1,a'} \ \forall a' \in S_1'$.

***Inductive step:*** For any time slot $t$, it is assumed that, $U_{t,a'}^{HLF-D} \geq U_{t,a'}^{\pi}, \forall a' \in S_t'$. It is required to prove that, $U_{t+1,a'}^{HLF-D} \geq U_{t+1,a'}^{\pi}, \forall a' \in S_{t+1}'$. Now, from the initial discussions,

$$S_t'^{HLF-D} = \{e_1, e_2, \dots, e_i, \dots, e_j, \dots, e_k, \dots, e_{R^{(t)}}\}_t^{HLF-D} \tag{21}$$

$$S_t'^{\pi} = \{e_1, e_2, \dots, e_i, \dots, e_j, \dots, e_k, \dots, e_{R^{(t)}}\}_t^{\pi}. \tag{22}$$

Now one of the four distinct cases may occur:

1. Critical samples are present in both $S_t'^{HLF-D}$ and $S_t'^{\pi}$.

2. Critical sample is present only in $S_t'^{HLF-D}$ but not in $S_t'^{\pi}$.

3. Critical sample is present only in $S_t'^{\pi}$ but not in $S_t'^{HLF-D}$.

4. Critical samples are not present in both $S_t'^{HLF-D}$ and $S_t'^{\pi}$.

Now, According to HLF-D, $d^{HLF-D(t)} = \arg \max\limits_{a' \in S_t'^{HLF-D}} \{Pri_{t,a'}^{HLF-D}\} = \{e_1\}_t^{HLF-D}$ whereas, in policy $\pi$ any element $d^{\pi(t)} = \{e_k\}_t^{\pi} \neq \{e_1\}_t^{\pi} = \arg \max\limits_{a' \in S_t'^{\pi}} \{Pri_{t,a'}^{\pi}\}$.

So, we can say that starting from the same initial condition, HLF-D always processes the critical sample with the hard deadline but that is not the case in $\pi$. So, no critical packet is dropped in HLF-D but it may get dropped in $\pi$ and the corresponding flow-line goes out of service. After reset, sensors from the temporarily inactive flow-lines join the active set again. Moreover, the numbers of new active sensors being added in each slot are equal. Hence, the number of elements present in $S_t^{\pi}$ is always equal or less than that in $S_t^{HLF-D}$. For comparing HLF-D and $\pi$, the number of elements in $S_t^{\pi}$ is made the same as that in $S_t^{HLF-D}$ by appending zeros at the beginning of $S_t^{\pi}$.

Now, we compare HLF-D and $\pi$ for the above mentioned four cases one by one.

## A. *Case 1:*

In the active sets $S_t'^{HLF-D}$, $S_t'^{\pi}$, $\{e_1\}_t^{HLF-D}$ and $\{e_1\}_t^{\pi}$ are critical, respectively. HLF-D always serves $\{e_1\}_t^{HLF-D}$ but $\pi$ serves any $0 < \{e_k\}_t^{\pi} \neq \{e_1\}_t^{\pi}$. So, the unserved critical sample $\{e_1\}_t^{\pi}$ is dropped from $S_t'^{\pi}$ at the end of the slot $t$. From the earlier discussions,

$$S_{t+1}^{HLF-D} = \left[ S_t'^{HLF-D} \backslash \{d^{HLF-D(t)}, drop_t^{HLF-D}\} \right] \cup N_{t+1} \cup RN_{t+1}^{HLF-D}. \tag{23}$$

In HLF-D, $drop_t^{HLF-D} = \{\emptyset\}$, $RN_{t+1}^{HLF-D} = \{\emptyset\}$, and $d^{HLF-D(t)} = \{e_1\}_t^{HLF-D}$ always. So,

$$S_t'^{HLF-D} = \{\cancel{e_1}, e_2 \dots e_k \dots e_{R^{(t)}}\}_t^{HLF-D}$$

$$S_{t+1}^{HLF-D} = \{e_2 \dots e_k \dots e_{R^{(t)}}\}_t^{HLF-D} \cup \{y_1 \dots y_{r^{(t+1)}}\}_{t+1}^{HLF-D}$$

$$= \{x_1 \dots x_{k-1} \dots x_{R^{(t)}-1}, x_{R^{(t)}} \dots x_{R^{(t+1)}}\}_{t+1}^{HLF-D}$$

$$= \{S_t'^{HLF-D} \backslash \{e_1\}_t^{HLF-D}\} \cup N_{t+1}. \tag{24}$$

From (24), the relation between $S_{t+1}^{HLF-D}$ and $S_t'^{HLF-D}$ can be obtained as

$$\{x_a\}_{t+1}^{HLF-D} = \{e_{a'+1}\}_t^{HLF-D} \quad for\ 1 \le a = a' < R^{(t)}$$

$$= \{y_n\}_{t+1} \qquad for\ n = 1\ to\ r^{(t+1)},\ R^{(t)} \le a \le R^{(t+1)}. \tag{25}$$

Here, $\{e_{a'+1}\}_t^{HLF-D}\ for\ 1 \le a = a' < R^{(t)}$ are all non-critical samples at slot $t$ and $\{y_n\}_{t+1}\ for\ n = 1\ to\ r^{(t+1)}$ has just arrived at slot $t+1$. So, the utilization of the elements in $S_{t+1}^{HLF-D}$ will be,

$$U_{t+1,a}^{HLF-D} = U_{t+1,\{a'+1\}_t}^{HLF-D} \quad for\ 1 \le a = a' < R^{(t)}$$

$$= 1 \qquad for\ R^{(t)} \le a \le R^{(t+1)}. \tag{26}$$

Here, $U_{t+1,\{a'\}_t}^{HLF-D}$ implies utilization of $a'$ element from $S_t'^{HLF-D}$ in the next slot $t+1$.

Similarly for $\pi$,

$$S_{t+1}^{\pi} = \left[ S_t'^{\pi} \backslash \{ d^{\pi(t)}, drop_t^{\pi} \} \right] \cup N_{t+1} \cup RN_{t+1}^{\pi}. \tag{27}$$

In $\pi$, $drop_t^{\pi} = \{e_1\}_t^{\pi}$ and $d^{\pi(t)} = \{e_k\}_t^{\pi} \ne \{e_1\}_t^{\pi}$. So,

$$S_t'^{\pi} = \{ \cancel{e_1}, e_2 \ldots e_{k-1}, \cancel{e_k}, e_{k+1} \ldots e_{R^{(t)}} \}_t^{\pi}$$

$$S_{t+1}^{\pi} = \{ 0, e_2 \ldots e_{k-1}, e_{k+1} \ldots e_{R^{(t)}} \}_t^{\pi} \cup \{ y_1 \ldots y_{r^{(t+1)}} \}_{t+1}^{\pi} \cup \{ \chi_1 \ldots \chi_{\varepsilon^{(t+1)}} \}_{t+1}^{\pi}$$

$$= \{ S_t'^{\pi} \backslash \{e_1, e_k\}_t^{\pi} \} \cup N_{t+1} \cup RN_{t+1}^{\pi}. \tag{28}$$

From the above equation, we can see that there are $\varepsilon^{\pi(t+1)}$ number of new sensor samples from regenerated flow-lines that are joining $S_{t+1}^{\pi}$ at the beginning of $t+1$: $RN_{t+1}^{\pi} = \{\chi_\eta\}_{t+1}^{\pi}\ for\ \eta = 1 \ldots \varepsilon^{\pi(t+1)}$ arranged in the decreasing order of the priority of the elements. These renewed samples are replacing 0s at the beginning of $S_{t+1}^{\pi}$ starting from the last 0's position. Let us consider that the last 0 is placed at $h^{th}$ position starting from the beginning of $S_{t+1}^{\pi}$ and $h - \varepsilon^{\pi(t+1)} + 1 = g$. Starting from $(h+1)^{th}$ position up to the end, all the elements have non-zero utility. So, the set $S_{t+1}^{\pi}$ is represented as follows,

$$S_{t+1}^{\pi} = \{ \overbrace{0, e_2, \ldots, e_{g-1}}^{All\ zeros}, \overbrace{0,0,\ldots,0}^{\chi_1, \chi_2 \ldots \chi_{\varepsilon^{\pi(t+1)}}}, e_{h+1}, \ldots, e_i \ldots e_j, \ldots, e_{k-1}, e_{k+1} \ldots, e_{R^{(t)}}, y_1, y_2, \ldots, y_{r^{(t+1)}} \}_t^{\pi}$$

$$= \{ x_1, x_2 \ldots x_{g-1}, x_g \ldots x_h, x_{h+1} \ldots x_i \ldots x_j \ldots x_{k-1}, x_k \ldots x_{R^{(t)}-1}, x_{R^{(t)}} \ldots x_{R^{(t+1)}} \}_t^{\pi}. \tag{29}$$

From (29), the relation between $S_{t+1}^{\pi}$ and $S_t'^{\pi}$,

$$\{x_a\}_{t+1}^{\pi} = 0 \qquad for\ 1 \le a = a' < g.$$

$$= \{\chi_\eta\}_{t+1}^{\pi} \qquad for\ \eta = 1 \ldots \varepsilon^{\pi(t+1)}, g \le a \le h.$$

$$= \{e_{a'}\}_t^{\pi} \qquad for\ h < a = a' < k.$$

$$= \{e_{a'+1}\}_t^{\pi} \qquad for\ \ k \le a = a' < R^{(t)}.$$

$$= \{y_n\}_{t+1} \qquad for\ n = 1\ to\ r^{(t+1)},\ R^{(t)} \le a \le R^{(t+1)}. \tag{30}$$

And the utility of the elements in $S_{t+1}^{\pi}$ will be,

$$\begin{aligned}
U_{t+1,a}^{\pi} &= 0 && for\ 1 \le a = a' < g. \\
&= U_{t+1,\chi_\eta}^{\pi} && for\ \eta = 1\dots\varepsilon^{\pi(t+1)}, g \le a \le h. \\
&= U_{t+1,\{a'\}_t}^{\pi} && for\ h < a = a' < k. \\
&= U_{t+1,\{a'+1\}_t}^{\pi} && for\ k \le a = a' < R^{(t)}. \\
&= 1 && for\ n = 1\ to\ r^{(t+1)},\ R^{(t)} \le a \le R^{(t+1)}.
\end{aligned} \tag{31}$$

Now comparing (26) and (31) it is observed that,

**for $1 \le a = a' < g$** : $U_{t+1,a}^{HLF-D} = U_{t+1,\{a'\}_t}^{HLF-D} \ge 0$ and $U_{t+1,a}^{\pi} = 0$. So, $U_{t+1,a}^{HLF-D} \ge U_{t+1,a}^{\pi}$.

**for $g \le a \le h$** : Any $\{\chi_\eta\}_{t+1}^{\pi}$ for $\eta = 1\dots\varepsilon^{\pi(t+1)}$ is a regenerated element. So, it has attempt $\xi_{\{\chi_\eta\}_{t+1}^{\pi}} \ge 2 \ \forall\ \{\chi_\eta\}_{t+1}^{\pi} \in RN_{t+1}^{\pi}$. But all the elements $\{x_a\}_{t+1}^{HLF-D}$ for $1 \le a \le R^{(t+1)}$ has attempt 1. So, $U_{t+1,a}^{HLF-D} = U_{t+1,\{a'\}_t}^{HLF-D} \ge U_{t+1,\chi_\eta}^{\pi} = U_{t+1,a}^{\pi}$.

**for $h < a = a' < k$** : All sensor samples are non-critical in this range in both $S_t'^{HLF-D}$ and $S_t'^{\pi}$ and they are arranged as follows,

$$Pri_{t,a'} > Pri_{t,a'+1}$$
$$or,\ U_{t,a'} \le U_{t,a'+1}. \tag{32}$$

Combining the initial assumption of the induction step and (32), it is obtained that $U_{t,a'+1}^{HLF-D} \ge U_{t,a'}^{\pi}$. Next, from Lemma-I, we get, $U_{t+1,\{a'+1\}_t}^{HLF-D} \ge U_{t+1,\{a'\}_t}^{\pi}$. Therefore, $U_{t+1,a}^{HLF-D} \ge U_{t+1,a}^{\pi}$.

**for $k \le a = a' < R^{(t)}$** : From assumption, $U_{t,a'+1}^{HLF-D} \ge U_{t,a'+1}^{\pi}$. So from Lemma-I, $U_{t+1,\{a'+1\}_t}^{HLF-D} \ge U_{t+1,\{a'+1\}_t}^{\pi}$. That means, $U_{t+1,a}^{HLF-D} \ge U_{t+1,a}^{\pi}$ for $k \le a = a' < R^{(t)}$.

**for $R^{(t)} \le a \le R^{(t+1)}$** : $U_{t+1,a}^{HLF-D} = U_{t+1,a}^{\pi} = 1$. So, $U_{t+1,a}^{HLF-D} \ge U_{t+1,a}^{\pi}$.

From the above comparisons, it can be concluded that if $U_{t,a'}^{HLF-D} \ge U_{t,a'}^{\pi} \forall a' \in S_t'$, $U_{t+1,a}^{HLF-D} \ge U_{t+1,a}^{\pi} \forall a \in S_{t+1}$ holds always. Again the similar four sub-cases may occur in slot $t+1$, depending on the presence of the critical sample in any active set, as mentioned at the beginning of the induction step. They are analyzed case by case to complete the induction.

*A.1. Subcase 1.1:* Let, $\{x_I\}_{t+1}^{HLF-D}$ and $\{x_J\}_{t+1}^{\pi}$ are critical in $S_{t+1}^{HLF-D}$ and $S_{t+1}^{\pi}$, respectively. So, $\{x_I\}_{t+1}^{HLF-D}$ and $\{x_J\}_{t+1}^{\pi}$ comes at the first place ($a' = 1$) in $S_{t+1}^{HLF-D}$ and $S_{t+1}^{\pi}$, respectively.

Now, the relations between sets $S_{t+1}^{HLF-D}$ and $S_{t+1}'^{HLF-D}$ are as follows,

$$\begin{aligned}
S_{t+1}^{HLF-D} &= \left\{ x_1, x_2 \dots x_{I-1}, x_I, x_{I+1} \dots x_{R^{(t+1)}} \right\}_{t+1}^{HLF-D} \\
S_{t+1}'^{HLF-D} &= \left\{ x_I, x_1, x_2 \dots x_{I-1}, x_{I+1} \dots x_{R^{(t+1)}} \right\}_{t+1}^{HLF-D} \\
&= \left\{ e_1, e_2, e_3 \dots e_I, e_{I+1}, \dots e_{R^{(t+1)}} \right\}_{t+1}^{HLF-D}.
\end{aligned} \tag{33}$$

This implies,

$$\begin{aligned}
\{e_{a'}\}_{t+1}^{HLF-D} &= \{x_I\}_{t+1}^{HLF-D} && for\ a' = 1 \\
&= \{x_{a-1}\}_{t+1}^{HLF-D} && for\ 1 < a = a' \le I \\
&= \{x_a\}_{t+1}^{HLF-D} && for\ I < a = a' \le R^{(t+1)}.
\end{aligned} \tag{34}$$

$$\begin{aligned}
U_{t+1,a'}^{HLF-D} &= U_{t+1,I}^{HLF-D} && for\ a' = 1 \\
&= U_{t+1,a-1}^{HLF-D} && for\ 1 < a = a' \le I
\end{aligned}$$

$$= U_{t+1,a}^{HLF-D} \quad for\ I < a = a' \le R^{(t+1)}. \tag{35}$$

Comparing $U_{t+1,a'}^{HLF-D}$ from (35) with $U_{t+1,a'}^{\pi}\ for\ a' = 1$ produces $U_{t+1,a'}^{HLF-D} = U_{t+1,I}^{HLF-D}$ and $U_{t+1,a'}^{\pi} = U_{t+1,J}^{\pi}$. It is known that starting from the same initial condition, HLF-D always serves critical sample or sample having the lowest utilization (in absence of critical sample). Other active samples with higher utilization are inherited as it is to the active set in the next slot. Whereas, $\pi$ serves sample with higher utilization other than the critical one and the unattended critical packet is dropped. As a result, one of the following two cases may occur :

i.   Critical sample present in $S_{t+1}^{\pi}$ at any slot $t + 1$ is same as that present in $S_{t+1}^{HLF-D}$.

ii.  Two different critical samples are present in $S_{t+1}^{HLF-D}$ and $S_{t+1}^{\pi}$. This is possible only under the circumstances as follows: Let us assume that the same sample having the highest latency is present in the active sets of both HLF-D and $\pi$ policy at any of the earlier slots. This sample is not served by $\pi$ but, it is served by HLF-D. This sample is then inherited or regenerated to the future active sets of $\pi$ and is present in $S_{t+1}^{\pi}$ as the critical sample. So, this critical sample in $S_{t+1}^{\pi}$ must have lower utilization than that of the first element $\{x_1\}_{t+1}^{HLFD}$ in $S_{t+1}^{HLF-D}$. Here, $U_{t+1,I}^{HLF-D} \ge U_{t+1,1}^{HLF-D} \ge U_{t+1,J}^{\pi}$.

Hence, $U_{t+1,J}^{\pi} \ge U_{t+1,I}^{HLF-D}$ or $U_{t+1,a'}^{HLF-D} \ge U_{t+1,a'}^{\pi}\ for\ a' = 1$.

Next, the regenerated elements $\chi_1 = x_g, \chi_2 = x_{g+1}, \ldots, \chi_{\pi^{(t+1)}} = x_h$ from $S_{t+1}^{\pi}$ are placed in $l_1, l_2, \ldots, l_{\varepsilon^{\pi(t+1)}}$ positions, respectively, in $S_{t+1}'^{\pi}$, according to the value of their priority. So, the set $S_{t+1}'^{\pi}$ becomes as follows,

$$S_{t+1}'^{\pi} = \left\{ \boxed{x_J}\ x_1, \ldots x_{g-1}, x_{h+1}, \ldots, x_g, \ldots x_{g+1} \ldots x_h, \ldots, x_{R^{(t+1)}} \right\}_t^{\pi}$$
$$= \left\{ \boxed{e_1}\ e_2, \ldots, e_{g-1}, e_g, \ldots e_{l_1}, \ldots e_{l_2}, \ldots, e_{l_{\pi(t+1)}}, \ldots e_{R^{(t+1)}} \right\}_t^{\pi} \tag{36}$$

In $S_{t+1}'^{\pi}$, all the elements except $e_1$ are non-critical and they are arranged in decreasing order of priority. So, all elements starting from $e_2$ to $e_{l_{\pi(t+1)-1}}$ have equal or lesser utilization than $e_{l_{\pi(t+1)}}$. Again, $\xi_{e_{l_{\pi(t+1)}}} \ge 2$ whereas all the elements $\{e_{a'}\}_{t+1}^{HLF-D}\ for\ 1 \le a' \le R^{(t+1)}$ has attempt 1. Therefore, directly it can be concluded that up to $l_{\varepsilon^{\pi(t+1)}}^{th}$ position any element $\{e_{a'}\}_{t+1}$ in $\pi$ has lower utility than $a'^{th}$ element in HLF-D : $U_{t+1,a'}^{HLF-D} \ge U_{t+1,a'}^{\pi}\ for\ 1 < a' \le l_{\varepsilon^{\pi(t+1)}}$. Now, to compare the utility of the rest of the elements in HLF-D and $\pi$, the knowledge about the position of the last renewed element $e_{l_{\pi(t+1)}}$ is required.

i.   If $g \le l_{\varepsilon^{\pi(t+1)}} \le J$ :

$$\{e_{a'}\}_{t+1}^{\pi} = \{x_{a-1}\}_{t+1}^{\pi} \quad for\ l_{\varepsilon^{\pi(t+1)}} < a = a' \le J$$
$$= \{x_a\}_{t+1}^{\pi} \quad for\ J < a = a' \le R^{(t+1)}. \tag{37}$$

$$U_{t+1,a'}^{\pi} = U_{t+1,a-1}^{\pi} \quad for\ l_{\varepsilon^{\pi(t+1)}} < a = a' \le J$$
$$= U_{t+1,a}^{\pi} \quad for\ J < a = a' \le R^{(t+1)}. \tag{38}$$

ii.  If $J < l_{\varepsilon^{\pi(t+1)}} \le R^{(t+1)}$ :

$$\{e_{a'}\}_{t+1}^{\pi} = \{x_a\}_{t+1}^{\pi} \quad for\ l_{\varepsilon^{\pi(t+1)}} < a = a' \le R^{(t+1)} \tag{39}$$

$$U_{t+1,a'}^{\pi} = U_{t+1,a}^{\pi} \quad for\ l_{\varepsilon^{\pi(t+1)}} < a = a' \le R^{(t+1)}. \tag{40}$$

Now, comparing $U_{t+1,a'}^{\pi}$ from (38) or (40) with $U_{t+1,a'}^{HLF-D}$ from (35) for $a' > l_{\varepsilon^{\pi(t+1)}}$ where, $l_{\varepsilon^{\pi(t+1)}}$ can take any value, we can directly conclude that, $U_{t+1,a'}^{HLF-D} \ge U_{t+1,a'}^{\pi}$ if $I \le J$. For $I > J$, $U_{t+1,a'}^{HLF-D} \ge U_{t+1,a'}^{\pi}$ is obtained $if\ f\ I < k$ from backtracking (26) and (31) by (35) and (38) or (40), respectively. So, our next aim is to show that $I$ is always less than $k$ under any circumstances.

The two algorithms HLF-D and $\pi$ start from the same initial condition. So their elements are the same at the beginning of $t = 1$. HLF-D always serves the critical sample or the sample with the highest priority when no critical sample is present which is not the case in $\pi$. So packets may drop in $\pi$. If the same critical sample be present both in $S_{t+1}^{HLF-D}$ and $S_{t+1}^{\pi}$ and its position be $I$ in $S_{t+1}^{HLF-D}$, then in $S_{t+1}^{\pi}$, it will be in the same position or after $I$ (due to initial 0 padding). That means $I \leq J$. But $I > J$ means two completely different samples are critical in $S_{t+1}^{HLF-D}$ and $S_{t+1}^{\pi}$. This can happen only when the element from $S_{t+1}^{\pi}$ same as the $I^{th}$ sample in $S_{t+1}^{HLF-D}$ is already served by $\pi$ at any earlier slot, i.e. $t$ or before. Let say, the $k^{th}$ sample in $S'^{\pi}_{t}$ was same as the current $I^{th}$ sample in $S_{t+1}^{HLF-D}$. Now, at any slot $t$, the first element from $S'^{HLF-D}_{t}$ is being served and other elements just shift one position left and inherits to the active set $S_{t+1}^{HLF-D}$ in the next slot. So, $I^{th}$ sample in $S_{t+1}^{HLF-D}$ was in $(I+1)^{th}$ position in $S'^{HLF-D}_{t}$ and earlier in this paragraph we discussed that if the same element is present in both $S_{t+1}^{\pi}$ and $S'^{\pi}_{t+1}$, they are placed either at the similar position or its position in $S'^{\pi}_{t+1}$ is after than that in $S_{t+1}^{\pi}$. Therefore, between $S'^{\pi}_{t}$ and $S'^{HLF-D}_{t}$, $k \geq (I+1)$, or in other words $I$ is always less than $k$.

**NOTE I.** In case 1, dropping of an unserved critical packet has occurred for $\pi$ policy. So, at least one 0 is there at the beginning of $S_{t+1}^{\pi}$ and $l_{\varepsilon^{\pi(t+1)}} \geq 1$. So, here the comparison of $S'^{HLF-D}_{t+1}$ and $S'^{\pi}_{t+1}$ always starts from their position $a' > 1$ depending on the value of $l_{\varepsilon\pi(t+1)}$.

**NOTE II.** $l_{\varepsilon\pi(t+1)}$ in $S'^{\pi}_{t+1}$ is the last position of the regenerated elements from $S_{t+1}^{\pi}$. After arranging $S_{t+1}^{\pi}$ in decreasing order of priority, the positions of $\chi_{\eta}$ can remain unchanged or anywhere after $g$ but the last position $l_{\varepsilon\pi(t+1)}$ of the regenerated elements can not be placed anywhere before $h$ as $g + \varepsilon^{\pi(t+1)} - 1 = h$.

**A.2. Subcase 1.2:** Critical sample present only in $S_{t+1}^{HLF-D}$ but not in $S_{t+1}^{\pi}$. Let, $\{x_I\}_{t+1}^{HLF-D}$ be the critical in $S_{t+1}^{HLF-D}$. Critical sample always have the highest priority. So, in $S'^{HLF-D}_{t+1}$, $\{x_I\}_{t+1}^{HLF-D}$ from $S_{t+1}^{HLF-D}$ inherits to $\{e_1\}_{t+1}^{HLF-D}$ positions and the relations between $S_{t+1}^{HLF-D}$ and $S'^{HLF-D}_{t+1}$ will be same as (34) and (35). $S'^{\pi}_{t+1}$ will be as follows:

$$S'^{\pi}_{t+1} = \left\{ x_1, x_2, \dots x_{g-1}, x_{h+1}, \dots x_g, \dots x_{g+1}, \dots x_h, \dots x_{R^{(t+1)}} \right\}^{\pi}_{t}$$
$$= \left\{ e_1, e_2 \dots e_{g-1}, e_g \dots e_{l_1} \dots e_{l_2} \dots e_{l_{\varepsilon^{\pi(t+1)}}} \dots e_{R^{(t+1)}} \right\}^{\pi}_{t} \tag{41}$$

From (31) we know that, $U_{t+1,a}^{\pi} = 0$ for $1 \leq a = a' < g$. Moreover, from the discussion in *subcase 1.1* we know that up to $l_{\varepsilon^{\pi(t+1)}}^{th}$ position in $S'^{\pi}_{t+1}$, any element $\{e_{a'}\}_{t+1}^{\pi}$ has lower utility than $a'$ element in $S'^{HLF-D}_{t+1}$. So, $U_{t+1,a'}^{HLF-D} \geq U_{t+1,a'}^{\pi}$ for $1 \leq a = a' \leq l_{\varepsilon^{\pi(t+1)}}$. Now, for the rest of the elements, the relations between $S_{t+1}^{\prime\pi}$ and $S_{t+1}^{\pi}$ are as follows:

$$\left. \begin{array}{l} \{e_{a'}\}_{t+1}^{\pi} = \{x_a\}_{t+1}^{\pi} \\ U_{t+1,a'}^{\pi} = U_{t+1,a}^{\pi} \end{array} \right\} \text{ for } l_{\varepsilon^{\pi(t+1)}} < a = a' \leq R^{(t+1)}. \tag{42}$$

Now, to compare (35) and (42), we backtrack (26) and (31), respectively and get,

$$U_{t+1,a'}^{HLF-D} = U_{t+1,\{a'\}_t}^{HLF-D} \qquad \text{for } l_{\varepsilon^{\pi(t+1)}} < a = a' \leq I$$
$$= U_{t+1,\{a'+1\}_t}^{HLF-D} \qquad \text{for } I < a = a' < R^{(t)}$$
$$= 1 \qquad \text{for } R^{(t)} \leq a \leq R^{(t+1)}. \tag{43}$$

$$U_{t+1,a'}^{\pi} = U_{t+1,\{a'\}_t}^{\pi} \qquad \text{for } h < a = a' < k$$
$$= U_{t+1,\{a'+1\}_t}^{\pi} \qquad \text{for } k \leq a = a' < R^{(t)}$$
$$= 1 \qquad \text{for } R^{(t)} \leq a \leq R^{(t+1)}. \tag{44}$$

From (43) and (44), it can be concluded that, $U_{t+1,a'}^{HLF-D} \geq U_{t+1,a'}^{\pi}$ iff $I < k$. Now, from *subcase 1.1* we know that $I$ is always less than $k$.

**A.3. Subcase 1.3:** Critical sample present only in $S_{t+1}^{\pi}$ but not in $S_{t+1}^{HLF-D}$. Let, $\{x_J\}_{t+1}^{\pi}$ be the critical in $S_{t+1}^{\pi}$. Critical sample always has the highest priority. So, in $S_{t+1}^{\prime\pi}$, $\{x_J\}_{t+1}^{\pi}$ from $S_{t+1}^{\pi}$ inherits to $\{e_1\}_{t+1}^{\pi}$ position and $S_{t+1}^{\prime\pi}$ is the same as (36). $S'^{HLFD}_{t+1}$ remains as it is $S_{t+1}^{HLFD}$. Now, comparing $S'^{HLFD}_{t+1}$ and $S'^{\pi}_{t+1}$ we get, $U_{t+1,1}^{HLF-D} \geq U_{t+1,J}^{\pi}$ and $U_{t+1,a'}^{HLF-D} \geq U_{t+1,a'}^{\pi}$ for $1 < a = a' \leq$ $l_{e\pi(t+1)}$ (from *subcase 1.1*). For the rest of the elements, $U_{t+1,a'}^{HLF-D}$ is compared to $U_{t+1,a'}^{\pi}$ from (38) or (40) with the help of (26) and (31), respectively. From that, it can be concluded that, $U_{t+1,a'}^{HLF-D} \geq U_{t+1,a'}^{\pi}$ for any value of $J$ and $k$.

***A.4. Subcase 1.4:*** Critical sample is not present both $S_{t+1}^{HLF-D}$ and $S_{t+1}^{\pi}$. So, $S_t^{\prime HLFD}$ and $S_t^{\prime \pi}$ remain the same as $S_{t+1}^{HLFD}$ and $S_{t+1}^{\pi}$ respectively. Now, for $S_{t+1}^{HLFD}$ and $S_{t+1}^{\pi}$ it is already proved that $U_{t+1,a}^{HLF-D} \geq U_{t+1,a}^{\pi}$ in case 1.

From the above comparisons in subcases [1.1-1.4], it can be concluded that $U_{t+1,a'}^{HLF-D} \geq U_{t+1,a'}^{\pi}$, $\forall a' \in S_{t+1}'$ always for case 1.

## B. Case 2:

In $S_t^{\prime HLF-D}$, $\{e_1\}_t^{HLF-D}$ is critical with the highest priority. $S_t^{\prime \pi}$ is same as $S_t^{\pi}$. HLF-D always serves $\{e_1\}_t^{HLF-D}$ but $\pi$ serves any $\{e_k\}_t^{\pi} > 0$ and $\{e_k\}_t^{\pi} \neq \{e_1\}_t^{\pi}$. Here no sample drops from $S_t^{\pi}$. Next, the relation between $S_{t+1}^{HLF-D}$ and $S_t^{\prime HLF-D}$ will be same as (25) & (26) whereas that of $S_{t+1}^{\pi}$ and $S_t^{\prime \pi}$ are found similar to (30) & (31).

We have already compared (26) and (31) in *case 1* of this induction and proved that, $U_{t+1,a}^{HLF-D} \geq U_{t+1,a}^{\pi}$ $\forall a \in S_{t+1}$. Again, the similar four sub-cases may occur in slot $t + 1$, depending on the presence of the critical sample in any active set, as mentioned at the beginning of the induction. They are analyzed case by case (*subcase 1.1-1.4*) similar to those in *case 1* to complete the induction step.

## C. Case 3:

In $S_t^{\prime \pi}$, $\{e_1\}_t^{\pi}$ is critical but $\pi$ serves any $\{e_k\}_t^{\pi} > 0$ and $\{e_k\}_t^{\pi} \neq \{e_1\}_t^{\pi}$. So, $\{e_1\}_t^{\pi}$ is dropped in $S_{t+1}^{\pi}$. The relation between $S_{t+1}^{\pi}$ and $S_t^{\prime \pi}$ will be the same as (30) & (31). $S_t^{\prime HLF-D}$ stays same as $S_t^{HLFD}$. Next, the proof is the same as *case 1*, and it can be shown that $U_{t+1,a'}^{HLF-D} \geq U_{t+1,a'}^{\pi}$ $\forall a' \in S_{t+1}'$ for *case 3*.

## D. Case 4:

Though critical samples are not present in both in $S_t^{\prime HLF-D}$ and $S_t^{\prime \pi}$, still, HLF-D serves $\{e_1\}_t^{HLF-D}$ but $\pi$ serves any $\{e_k\}_t^{\pi} > 0$ and $\{e_k\}_t^{\pi} \neq \{e_1\}_t^{\pi}$. In this case, no sample drops from $S_t^{\prime \pi}$. Next, similar to *case 2* it can be proved that $U_{t+1,a'}^{HLF-D} \geq U_{t+1,a'}^{\pi}$ $\forall a' \in S_{t+1}'$ for *case 4*.

So, from the above discussions of all four cases and their subcases, it can be concluded that if $U_{t,a'}^{HLF-D} \geq U_{t,a'}^{\pi}$ $\forall a' \in S_t'$ at any slot $t$, in the next slot $t + 1$, $U_{t+1,a'}^{HLF-D} \geq U_{t+1,a'}^{\pi}$ $\forall a' \in S_{t+1}'$ as well **(Induction complete).** ∎

# V. Results And Discussion

This section plots the results obtained from MATLAB simulations indicating the performance evaluation of HLF-D algorithm in terms of the expected weighted sum utilization (EXWSUoI), age, latency and jitter (Here, EXWSUoI has the unit (No of time slots)⁻¹ . Rest all the parameters are expressed in terms of the number of time slots) for scheduling real-time updates with deadlines in symmetric IWSAN network with flow-line regenerations. Its results are then compared with those of other scheduling algorithms like newly introduced Highest Latency First (HLF) policy [34], traditional Earliest Deadline First (EDF), and Least Laxity First (LLF) algorithms. According to EDF, a sample with the earliest absolute deadline has the highest priority than others. LLF assigns the highest priority to the sample that currently having the lowest laxity/slack time left. HLF schedules a packet with the highest amount of latency involved so far.

However, for all the simulations, parameters used and their values are listed below in Table-II. Initial age is considered to be $h_{1,i}(1) \leq c_i(0) + D_i(1)$ slots.

Fig. 5. plots the comparison of EXWSUoI values varying with T for different algorithms. From this plot, it can be seen that the expected utilization of HLF-D for any value of T.

Fig. 6. compares the mean age, latency and RMS jitter of the samples. It can be clearly seen that our proposed HLF-D algorithm provides the lowest values for all of the above-mentioned metrics than those of the other scheduling policies.

Though smaller age value should be maintained to keep the information fresh that helps in accurate decision making of the cyber systems, [34] proves that minimizing the latency content involved in the age is more effective for the thoughtful trade-off between the information freshness maximization and jitter minimization in CPSs. This can be done using HLF algorithm. Moreover, ICPS is a time-critical real-time system. Here information exchange should be done as timely as possible to guarantee real-time responses within specified time constraints often referred to as 'deadlines'. Whenever a packet misses its deadline, its time-to-live (TTL) [41] becomes zero and the packet is dropped causing the breakdown of the corresponding flow-line, production loss and/or even more fatal consequences. Therefore, minimizing packet loss has the utmost importance, too, in addition to the end-to-end delay minimization. But, HLF algorithm fails to achieve this performance goal. There needs another algorithm HLF-D [38] to overcome the limitations of HLF. This scheduling policy maximizes the expected utility value of all the information contents present in the sensor samples by minimizing their mean latency in addition to the minimum packet drops. So, one may aptly conclude that HLF-D is a more suitable deadline-aware scheduling strategy for optimizing the freshness of information and QoS in CPSs.

However, if any packet, unfortunately, misses its deadline due to resource constraints or bad channel conditions, there should be a provision of urgent fixing or replacement of flow-lines to avoid congestion and production loss in the system network. This work extends the applicability of HLF-D in deadline-aware scheduling of real-time updates for maximizing the freshness of information with regenerations of the broken flow-lines in ICPSs. Moreover, if we compare the expected utilization obtained from this work with that obtained in [38] it is found that the HLF-D and LLF perform more or less in the same way for both the cases whereas the performances of HLF and EDF have been improved significantly due to the flow-line regenerations in deadline-aware scheduling.

TABLE-II
SIMULATION PARAMETERS

| Parameters | Symbol | Value Assigned |
|---|---|---|
| Number of flow-line (sensor-actuator pair) [34], [40] | M | 16 |
| The weight assigned to each sensor | $\alpha_i = \alpha$ $\forall i \in M$ | 1 |
| Time-shared transmission channel reliability | $p_i = p$ $\forall i \in M$ | 0.8 |
| Time slot (According to WirelessHart standard [24]) | $t$ | 10 ms |
| Initial useful age [40] | $\overrightarrow{c(0)}$ | [1,25] |
| Initial deadline [41] | $\overrightarrow{D(1)}$ | [1,20] |
| Attempt Number | $\xi_i$ $\forall i \in M$ | 1 |
| Real Indices | $\beta , \gamma$ | 1 |

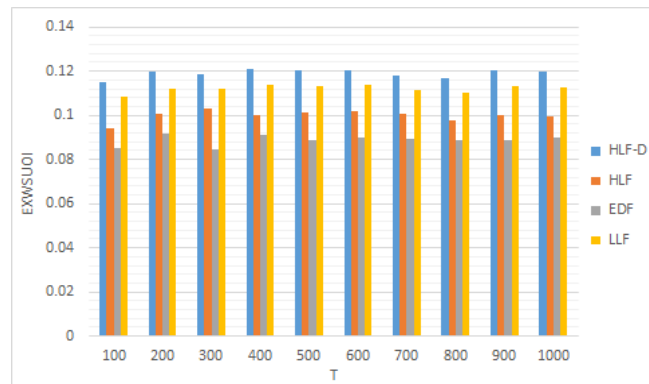

Fig. 5. Comparison of EXWSU for different algorithms.

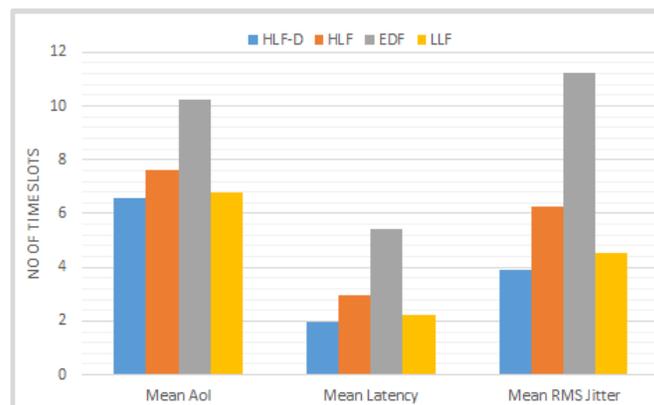

Fig. 6. Comparison of mean age, latency, and jitter for different algorithms.

## VI. CONCLUSION

This work analyzes the effectiveness of HLF-D algorithm for scheduling deadline-aware real-time updates with the help of flow-line regeneration mechanism in symmetric IWSAN for CPPS application. Here, one of the suitable active sensors is selected by the HLF-D policy to use an unreliable TDMA shared channel for sending real-time status update about the state of interest of the underlying system to a centralized processor or controller. Controller triggers appropriate control commands to the corresponding actuators and the actuators acts accordingly. If any packet fails to update its information to the controller within its associated deadline, the corresponding flow-line or CPS link breaks down. But, it joins the closed-loop CPS network again, once the problem is fixed. From mathematical analysis and simulations, it can be proved from this work that HLF-D outperforms other scheduling policies in preserving the usefulness of the information. Moreover, their effects on system performances (age, latency, and jitter) are also compared via simulations and from there it can be concluded that to guarantee a real-time response, data freshness and QoS simultaneously, our proposed algorithm HLF-D is the most effective packet scheduling scheme for ICPSs. Some interesting research advances of this work include the effect of multi-server, multi-path or erroneous channels in finding the age optimal scheduling algorithm for IWSANs.

## REFERENCES

[1]  R. Rajkumar, I. Lee, L. Sha, and J. Stankovic, "Cyber-Physical Systems: The Next Computing Revolution," in *Proc. 47th ACM/IEEE Design Automation Conf. (DAC)*, USA, June 13-18, 2010.

[2]  S. Kaul, R. Yates, and M. Gruteser, "Real-time status: How often should one update?", in *Proc. IEEE International Conference on Computer Communications (INFOCOM)*, USA, Mar 25-30, 2012.

[3]  M. Costa, M. Codreanu, and A. Ephremides, "On the age of information in status update systems with packet management," *IEEE Transactions on Information Theory*, vol. 62, no. 4, pp. 1897–1910, 2016.

[4]  I. Kadota, A. Sinha, and E. Modiano, "Scheduling Policies for Minimizing Age of Information in Broadcast Wireless Networks," submitted to *IEEE/ACM Transactions on Networking*. https://arxiv.org/abs/1801.01803 (Accessed on 31.05.2018).

[5]  A. Kosta, N. Pappas, and V. Angelakis, "Age of Information: A New Concept, Metric, and Tool," *Foundations and Trends in Networking*, vol. 12, no. 3, pp 162–259, 2017.

[6]  X. Wang and M. D. Lemmon, "Event-triggering in distributed networked control systems," *IEEE Transactions on Automatic Control*, vol. 56, no. 3, pp. 586-601, March 2011.

[7]  A. Molin and S. Hirche, "Event-Triggered State Estimation: An Iterative Algorithm and Optimality Properties," *IEEE Transactions on Automatic Control*, vol. 62, no. 11, pp. 5939-5946, November 2017.

[8]  M. J. Khojasteh, P. Tallapragada, J. Cortés, and Massimo Franceschetti "Time triggering vs. event triggering control over communication channels," *IEEE 56th Annual Conference on Decision and Control (CDC)*, Australia, December 12-15, 2017.

[9]  A. Cervin, M. Velasco, P. Marti, and A. Camacho, "Optimal online sampling period assignment: Theory and experiments," IEEE Transactions on Control Systems Technology, vol. 19, no. 4, pp. 902–910, Jul. 2011.

[10]  Y. Sun, Y. Polyanskiy, and E. Uysal-Biyikoglu, "Remote estimation of the Wiener process over a channel with random delay," *IEEE International Symposium on Information Theory (ISIT)*, Germany, June 25-30, 2017.

[11]  M. Costa, M. Codreanu, and A. Ephremides, "On the Age of Information in Status Update Systems with Packet Management," *IEEE Transactions on Information Theory*, vol. 62, no. 4, April 2016. pp. 1897-1910.

[12]  M. Rabi, G. V. Moustakides, and J. S. Baras, "Multiple sampling for estimation on a finite horizon," in *Proc. 45th IEEE Annual Conference on Decision and Control (CDC)*, USA, December 13-15, 2006.

[13]  P.R.Kumar and C. L. Robinson, "Sending the most recent observation is not optimal in networked control: Linear temporal coding and towards the design of a control specific transport protocol, *46th IEEE Conference on Decision and Control (CDC)*, USA, December 12-14, 2007.

[14]  D. Maity, M. H. Mamduhi , S. Hirche , K. H. Johansson, and J. S. Baras, "Optimal LQG Control Under Delay-Dependent Costly Information," IEEE Control Systems Letters, vol.3, no. 1, pp. 102-107, January 2019.

[15]  V. Gupta, A. F. Dana, J. P. Hespanha, R. M. Murray, and B. Hassibi, "Data Transmission Over Networks for Estimation and Control," *IEEE Transactions on Automatic Control*, vol. 54, no. 8, pp. 1807-1819, August 2009.

[16]  T. C. Koopmans, "Concepts of Optimality And Their Uses," Nobel Memorial Lecture, December 11, 1975.

[17]  M. H. Mamduhi, A. Molin, D. Tolic, and S. Hirche, "Error-dependent data scheduling in resource-aware multi-loop networked control systems," *Automatica*, vol. 81, pp. 209–216, Jul. 2017.

[18]  K. Huang, W. Liu†, Y. Li, and B. Vucetic, "To Retransmit or Not: Real-Time Remote Estimation in Wireless Networked Control", *IEEE International Conference on Communications (ICC)*, China, May 20-14, 2019.

[19]  Y. Li, F. Zhang, D. E. Quevedo, V. Lau, S. Dey, and L. Shi, "Power Control of an Energy Harvesting Sensor for Remote State Estimation," *IEEE Transactions on Automatic Control*, vol. 62, no. 7, pp. 277-290, January 20017.

[20]  A. S. Leong, D. E. Quevedo, D. Dolz, and S. Dey "Transmission Scheduling for Remote State Estimation Over Packet Dropping Links in the Presence of an Eavesdropper," *IEEE Transactions on Automatic Control*, vol. 64, no. 9, pp. 3732-3739, September 2019.

[21]  K. R. Baker and D. Trietsch, "Principles of Sequencing and Scheduling," John Wiley & Sons, Inc., 2009.

[22]  S. Wang, J. Wan, D. Li, and C. Zhang, "Implementing Smart Factory of Industrie 4.0: An Outlook," *Hindawi Publishing Corporation International Journal of Distributed Sensor Networks*, vol. Jan, 2016.

[23]  C. Lu, A. Saifullah, M. Sha, H. Gonzalez, D. Gunatilaka, C. Wu, L. Nie, and Y. Chen., "Real-Time Wireless Sensor-Actuator Networks for Industrial Cyber-Physical Systems," in *Proc. of the IEEE*, vol. 104, no. 5, pp. 1013 – 1024, 2016.

[24]  J. Åkerberg, M. Gidlund and M. Björkman, "*Future research challenges in wireless sensor and actuator networks targeting industrial automation,*" *IEEE 9th International Conference on Industrial Informatics (INDIN)*, Portugal, July 26-29, 2011.

[25]  C. Kam, S. Kompella, G. D. Nguyen, J. E. Wieselthier, and A. Ephremides, "Controlling the Age of Information: Buffer Size, Deadline, and Packet Replacement," *IEEE Military Communications Conference (MILCOM)*, USA, Nov 1-3, 2016.

[26]  C. Kam, S. Kompella, G. D. Nguyen, J. E. Wieselthier, and A. Ephremides, "On the Age of Information With Packet Deadline," *IEEE Transactions on Information Theory*, vol. 64, no. 9, pp. 6419 - 6428, 2018.

[27]  Y. Inoue, "Analysis of the Age of Information with Packet Deadline and Infinite Buffer Capacity," *IEEE International Symposium on Information Theory (ISIT)*, USA, June 17-22, 2018.

[28] Q. He, D. Yuan, and A. Ephremides, "Optimizing freshness of information: On minimum age link scheduling in wireless systems," in *Proceedings of 14th IEEE International Symposium on Modeling and Optimization in Mobile, Ad Hoc, and Wireless Networks (WiOpt)*, USA, May 9-13, 2016.

[29] A. M. Bedewy, Y. Sun, and N. B. Shroff, "Optimizing Data Freshness, Throughput, and Delay in Multi-Server Information-Update Systems," in *Proceedings of IEEE International Symposium on Information Theory (ISIT)*, Spain, July 10-15, 2017.

[30] R. Talak, S. Karaman, and E. Modiano, "Optimizing Information Freshness in Wireless Networks under General Interference Constraints," *Proceedings of the Eighteenth ACM International Symposium on Mobile Ad Hoc Networking and Computing (Mobihoc)*, USA, June 26 – 29, 2018.

[31] R. Talak, S. Karaman, and E. Modiano, "Distributed Scheduling Algorithms for Optimizing Information Freshness in Wireless Networks," *IEEE 19th International Workshop on Signal Processing Advances in Wireless Communications (SPAWC)*, Greece, 25-28 June, 2018.

[32] C. Joo and A. Eryilmaz, "Wireless Scheduling for Information Freshness and Synchrony: Drift-Based Design and Heavy-Traffic Analysis," *IEEE/ACM Transactions on Networking*, vol. 26, no. 6, pp. 2556 – 2568, December 2018.

[33] Z. Wang, X. Qin, B. Liu, and P. Zhang, "Joint Data Sampling and Link Scheduling for Age Minimization in Multihop Cyber-Physical Systems," *IEEE Wireless Communication Letters*, vol. 8, no. 3, pp. 765-768, January 2019.

[34] D. Sinha and R. Roy, "Scheduling Status Update for Optimizing Age of Information in the Context of Industrial Cyber-Physical System," *IEEE Access*, vol. 7, pp. 95677-95695, May 2019.

[35] R. Singh, X. Guo, and P.R. Kumar, "Index Policies for Optimal Mean-Variance Trade-Off of Inter-delivery Times in Real-Time Sensor Networks," *IEEE Conference on Computer Communications (INFOCOM)*, Hong Kong, April 26-May 1, 2015.

[36] X. Guo, R. Singh, and P.R. Kumar "A Risk-Sensitive Approach for Packet Inter-Delivery Time Optimization in Networked Cyber-Physical Systems," *IEEE/ACM Transactions on Networking*, vol. 26, no. 4, pp. 1976 – 1989, August 2018.

[37] N. Lu, B. Ji, and B. Li, "Age-based Scheduling: Improving Data Freshness for Wireless Real-Time Traffic," *Proceedings of the Eighteenth ACM International Symposium on Mobile Ad Hoc Networking and Computing (Mobihoc)*, USA, June 26 – 29, 2018.

[38] D. Sinha and R. Roy, "Deadline-aware Scheduling for Maximizing Freshness of Information in the Context of Industrial Cyber-Physical System," https://arxiv.org/abs/1912.11310 (access on 09.01.2020).

[39] F. Pérez, E. Irisarri, D. Orive, M. Marcos, and E. Estevez, "A CPPS Architecture approach for Industry 4.0," *IEEE 20th Conference on Emerging Technologies & Factory Automation (ETFA)*, Luxembourg, Sept 8-11, 2015.

[40] J. Åkerberg, M. Gidlund, J. Neander, T. Lennvall, and M. Björkman, "Deterministic Downlink Transmission in WirelessHART Networks enabling Wireless Control Applications," *36th Annual Conference on IEEE Industrial Electronics Society (IECON)*, USA, Nov 7-10, 2010.

[41] X. Wu, "Modeling and Performance Evaluation of Wirelesshart," Master's Thesis, University of Twente, Aug 31, 2012.

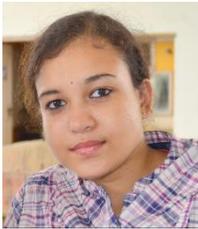
**Devarpita Sinha** received her B. Tech. degree in Electronics and Communication Engineering from West Bengal University of Technology, Kolkata, India, in 2014 and M.E. degree with First Class (Distinction) in the specialization of Wireless Communication from the Department of Electronics and Communication Engineering, Birla Institute of Technology (BIT) Mesra, Ranchi in 2016. She is currently pursuing the Ph.D. degree in the Department of Electronics and Electrical Communication Engineering, Indian Institute of Technology (IIT), Kharagpur, India. Her research interests include resource allocation problems, performance evaluation and control in communication networks, scheduling and decision making, etc. She is actively associated with IEEE Student branch, IIT Kharagpur and presently serving as the Chair of the same. She received prestigious Richard. E. Merwin Scholarship awarded by IEEE Computer Society in Spring Cycle, 2019 for her promising academic and leadership activities.

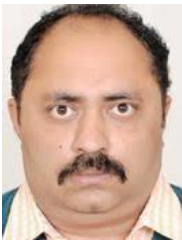
**Dr. Rajarshi Roy** received his B.E. degree with first-class (Hons.) in Electronics and Tele-Communication Engineering from Jadavpur University, Kolkata, West Bengal, India, in 1992, MSc. (Engineering) degree from the Department of Electrical Communication Engineering, IISc, Bangalore, India in 1995 and Ph.D. from Electrical Engineering Department, Polytechnic University, Brooklyn, New York, USA (Currently known as New York University, Tandon School of Engineering) in the year 2001. He is currently working as an Associate Professor in the Department of Electronics and Electrical Communication Engineering, Indian Institute of Technology, Kharagpur, India. His research interests include queuing theory, Markov decision theory, 5G/6G wireless communications, complex communication networks, MIMO, Network coding, Cooperative communication, cyber-physical systems, cognitive technology, etc.